\documentclass[11pt,a4paper]{article}
\usepackage{jheppub,bm,color}
\usepackage{yi}

\def \vJ {\bm{J}}

\def \tF {\tilde{F}}

\def \vpd {{\bm \pd}}
\def \vpdp {{\bm \pd}_{p}}
\def \vzeta {{\bm \zeta}}
\def \vg {{\bm g}}

\def \sO {{\cal O}}
\def \sT {{\cal T}}
\def \sA {{\cal A}}

\def \con {{\rm con}}

\begin{document}
\title{Towards a bottom-up formulation of spin kinetic theory}

\newcommand{\USTC}{Interdisciplinary Center for Theoretical Study, University of Science and Technology of China, Hefei, Anhui 230026, China}
\newcommand{\UIC}{Department of Physics, University of Illinois, Chicago, Illinois 60607, USA}
\newcommand{\LQT}{Laboratory for Quantum Theory at the Extremes, University of Illinois, Chicago, Illinois 60607, USA}

\author[a]{Zonglin Mo}
\affiliation[a]{\USTC}
\affiliation[a]{\UIC}

\author[b]{Yi Yin}
\affiliation[b]{
School of Science and Engineering, The Chinese University of Hong Kong (Shenzhen), Longgang, Shenzhen, Guangdong, 518172, China
}

\abstract{
We develop a bottom-up formulation of spin-kinetic theory for hot and/or dense plasmas. We introduce scalar and axial-vector phase-space functions as dynamical variables that parametrize both spin-averaged and spin-dependent distribution functions. Using spin-dependent Poisson brackets, we derive the corresponding kinetic equations and construct the associated Schwinger–Keldysh action. We further demonstrate how physical observables can be expressed in terms of these dynamical variables through constitutive relations. In the linear response regime, we establish a precise matching between the kinetic-theory and field-theory descriptions of vector and axial Wigner functions under electromagnetic and gravitational perturbations. Our framework provides a complementary approach to describing the dynamics of spin effects in a medium.
}

\emailAdd{zlmphy@mail.ustc.edu.cn}
\emailAdd{yiyin@cuhk.edu.cn}

\maketitle

\section{Introduction}

In this paper, we explore the "bottom-up" approach to formulating spin kinetic theory (SKT), which describes the dynamics of both spin-averaged and spin-dependent phase-space distributions in many-body systems, including those out of equilibrium. 
A prominent application of SKT is understanding spin-related phenomena in hot QCD matter, which are accessible through the polarization of Lambda hyperons~\cite{STAR:2017ckg,STAR:2019erd} (see Refs.~\cite{STAR:2022fan,Sheng:2022wsy} on phenomena for vector mesons) in heavy-ion collisions.
For example, vorticity-induced polarization~\cite{Liang:2004ph,Becattini:2013fla} along with other novel spin-related phenomena in QCD plasma, including shear-induced polarization~\cite{Liu:2021uhn,Fu:2021pok,Becattini:2021suc,Becattini:2021iol} and analogous spin Hall effect~\cite{Liu:2020dxg,Fu:2022myl}, are actively investigated both experimentally and theoretically~\cite {Becattini:2020ngo,Becattini2022,YinYi:2023msj}.

In response, significant developments have been made in formulating
the SKT for relativistic plasma~\cite{Hattori:2019ahi,Weickgenannt:2019dks,Gao:2019znl,Weickgenannt:2020aaf,Weickgenannt:2021cuo,Hayata2020,Sheng:2021kfc,Yang:2020hri} (see ref.~\cite{Hidaka:2022dmn} for a review).
A key predecessor to SKT is chiral kinetic theory, which describes the distribution of axial density/helicity~\cite{Son:2012wh,Stephanov:2012ki,Son:2012zy,Chen:2012ca,Gao:2012ix,Chen:2013iga,Manuel:2013zaa,Manuel:2014dza,Chen:2014cla,Chen:2015gta,Hidaka:2016yjf,Huang:2018wdl,Gao:2018wmr,Ma:2022ins}. 
Most of the references mentioned above, particularly those from recent years, derive SKT using what we shall refer to as the "top-down" approach. 
This approach begins with the microscopic Schwinger-Dyson equation for the Wigner function of the Fermionic field and then extracts the relevant equations for its components.

In contrast, the kinetic equation for the spin-averaged distribution can be constructed using a "bottom-up" strategy: one first proposes equations of motion for the distribution function, supplemented by constitutive relations for observables such as the phase-space vector current. 
This step is independent of microscopic details, which only enter as the next step when matching the kinetic theory's description to a microscopic calculation~\cite {lifshitz1981kinetics,Mueller:2002gd,Son:2012zy,Chen2016}. 
For example, kinetic theory for particle excitations in a hot gauge plasma reproduces the same induced current in a soft background gauge field as that obtained from the hard thermal loop effective theory~\cite{Blaizot:1993zk}. 
Perhaps a more familiar example is the correspondence between Landau Fermi liquid and diagrammatic analysis~\cite{abrikosov1975methods}.
This raises the question: can this bottom-up approach be extended to formulate SKT?

In this work, we propose and examine the following bottom-up construction of SKT.
We hope the resulting theory can be applied to systems such as a hot quark-gluon plasma in the weakly coupled regime.
Following the general procedure as outlined in sec.~\ref{sec:eff},
we propose a scalar and an axial vector function in phase space as dynamical variables to parametrize both the spin-averaged and spin-dependent phase-space distribution functions, and write down their E.o.M. using newly-formulated spin-dependent Poisson brackets (P.B.)~\cite{Delacretaz:2022ocm} in sec.~\ref{sec:spin-action}.
Also presented is the corresponding Schwinger-Keldysh action. 
We then express various observables, including vector and axial Wigner functions in phase space, in terms of dynamical variables through the constitutive relation in sec.~\ref{sec:constitutive}.
By comparing the response of kinetic theory as described in sec.~\ref{sec:kin-lin} with the lowest order linear response results around a homogeneous and static, but not necessarily equilibrium distribution in sec.~\ref{sec:field}, we show the precise matching between the two theories in sec.~\ref{sec:matching}.
This demonstrates that the "bottom-up" method is not only feasible, but also complements the existence approach.

\textbf{Notations}: We use $[\dots]_-$ and $[\dots]_+$ for commutator and anti-commutator of field operators to distinguish them from the Poisson bracket $\{\dots\}$ defined in eq.~\eqref{PB}. The convention for the Dirac matrix follows that in ref.~\cite{Ghiglieri:2020dpq} where $[\g^{\mu},\g^{\nu}]_+=-2g^{\mu\nu}$, 
which turns out to be convenient for the calculations with most plus metric $\eta^{\mu\nu}={\rm diag}(-1,1,1,1)$. 
We use the short-handed notation $\int_{p^{0}}=\int\frac{dp^{0}}{2\pi},\ \int_{\vp}=\int\frac{d^3\vp}{(2\pi)^3}$ for the integration in phase space. 
Fourier vector in Minkowski space is expressed as $Q^{\mu}=(\o,\vq)$.

\section{Formulation}

\subsection{General procedure
\label{sec:eff}
}

In this subsection, we outline the general steps of the bottom-up approach for establishing an effective description of dynamics in a many-body system (see also ref.~\cite {Liu:2018kfw}).
\begin{itemize}
\item   Step 1: Identify (propose) the relevant slow degrees of freedom (d.o.f.), denoted collectively by $\chi$;
\item Step 2: Construct the macroscopic equation of motion (E.o.M.) for $\chi$ 
\item Step 3: Express the observables of interest in terms of $\chi$, a relation known as the constitutive relation;

\item Step 4: Match the constitutive relation against microscopic calculations, particularly the non-analytic part of the response function. 
This step acts as a test for the effective description.
\end{itemize}

It is worth noting that the applicability of an effective approach typically relies on the existence of a clear separation between a soft scale (e.g., the characteristic gradient of the system ($q$) and a hard scale ($\Lambda_{H}$).
Consequently, the constitutive relation (see step 3) should be systematically constructed order by order using gradient expansion in the small parameter $q/\Lambda_{H}\sim \sO(\pd)$. 
At each order in gradient, one should include all possible terms consistent with the underlying physical principles, such as symmetry constraints. 
The coefficients in this expansion—the transport coefficients—are functionals of the slow variables $\chi$ and can be extracted by matching with microscopic calculations (i.e., through step 4).

For illustration, let us consider the description of a conserved $U(1)$ charge in the hydrodynamic limit. 
The charge density $J^{0}=\rho$ is identified as the slow variable (step 1) and obeys the conservation equation as its E.o.M. $\pd_{\mu}J^{\mu}=0$ (step 2).
The hard scale is the inverse of the mean free path $1/l_{\mfp}$ and the expansion parameter is $q l_{\mfp}\ll 1$. 
The constitutive relation for the spatial component of the current $\vJ$ can be expressed as follows (step 3)
\begin{align}
\label{j-Hy}
\vJ= -D\, \vpd\rho + \sigma(\rho) \vE +{\cal O}(\pd^{2})\, .
\end{align}
The first term is Fick's law, while the second one represents the Ohmic current.
Finally, this description should be contrasted with the linear response calculation in field theory (step 4).
Note, the associated response function in the Fourier space must be non-analytic in $\o, \vq$ due to the diffusive pole at $\o=- i D q^2$.

Before closing, let us mention that the effective description can be promoted to an effective field theory by constructing an effective action $I[\chi]$ for slow modes. 
The E.o.M.(s) correspond to the saddle point of the action. 
One advantage of the action formulation is that it makes the underlying symmetry transformation transparent—a feature that is less obvious at the level of the equations of motion.

In the following sections, we shall formulate SKT through the general steps outlined above.


\subsection{Spin kinetic theory, Poisson Brackets and Schwinger-Keldysh action
\label{sec:spin-action}
}

We assume that in the absence of perturbation/fluctuations, the system is homogeneous, though not necessarily in equilibrium. 
Our goal is to describe the subsequent evolution of observables, including those associated with spin, due to the presence of a disturbance.

\subsubsection{Dynamical variables}
We first propose dynamic variables, which consist of a scalar function \( f \) and an axial vector function \( \vg \). These functions will describe the spin-averaged and spin-dependent distribution functions in phase space, denoted as \( \G = (\vx, \vp) \). We assume that \( f \) has a non-zero background value \( f_0 \), while \( \vg \) vanishes. The latter reflects the assumption that the characterization of the background system does not depend on any pseudo-tensor quantities, such as axial charge density. 

It is important to note that the number of degrees of freedom in our approach differs from that in the top-down formulation, which starts with the equations governing all 16 components of Wigner functions in 3+1-dimensional Dirac theory. 
Thus, our construction is presented as a proposal that requires testing through microscopic calculations.

Next, we specify the regime in which the resulting kinetic theory is expected to apply. We assume that the typical frequency \( \omega \) and gradient \( q \) of the particle distribution satisfy the following conditions:
\begin{align}
    \label{eq:domain}
    \g_{R}, m_{D} \ll \omega, q \ll T_{\eff}\, . 
\end{align}
Here, \( \g_{R} \) represents the mean damping rate, \( m_{D} \) denotes the thermal mass, and the typical energy of the quasi-particle is referred to as the effective temperature \( T_{\eff} \). 

Because of the first inequality in eq.~\eqref{eq:domain}, we shall concentrate on the collisionless regime and neglect the self-energy corrections, which are of the order of \( m_{D} \). 
The second inequality ensures that a gradient expansion exists in powers of \( q/T_{\eff} \).
Note that this gradient expansion parameter differs from that in hydrodynamics, which is given by \( q/\gamma_{R} \). We focus on an effective description based on kinetic theory outside the hydrodynamic regime, consistent with our set-up that the background distribution \(f_0\) need not be in local thermal equilibrium.

Eq.~\eqref{eq:domain} assumes the separation of scales: \( m_{D}, \g_{R} \ll T_{\eff} \), which is true for a weakly coupled plasma with coupling constant \( g_C < 1 \) where \( m_{D} \sim g_{C} T_{\eff} \) and \( \g_{R} \sim g^{2}_{C} T_{\rm eff} \).

\subsubsection{Poisson bracket and equation of motion
\label{sec:EoM-kin}
}

The method we employ to write down the E.o.M. for $(f,\vg)$ is the elegant Poisson bracket technique, which was recently generalized to the spin-dependent case in ref.~\cite{Delacretaz:2022ocm}. 
Let us recall that the collisionless Boltzmann equation can be written in terms of Poisson brackets (P.B.) as
\begin{align}
\label{Boltzmann-f}
    \partial_{t}f  +\{f,\e\}=0\, ,
\end{align}
where for arbitrary function $a,b$ in the phase space, 
P.B. is defined as
\begin{align}
\label{PB}
    \{a, b\}(\Gamma) =
       (\vpd a)\cdot (\vpdp b)-(\vpdp a)\cdot(\vpd b)\, . 
\end{align}
Here, $\boldsymbol{\partial}_p$ denotes the derivative with respect to $\boldsymbol{p}$.
The single particle energy function(al) $\e$ can be written as
\begin{align}
    \e= \e_{\vp}+\delta \e\, .
\end{align}
Here, $\e_{\vp}$ is the background single particle energy and $\d\e$ represents the deviation.

We now recall the definition of spin-dependent P.B.~\cite{Delacretaz:2022ocm}.
Let us introduce "big vector" $E=(E_0,E_{i}), G=(G_0,G_{i})$ in the phase space and denote the generalized P.B. as $\{E,G\}=(\{E,G\}_0,\{E,G\}_i)$.
The latter is explicitly given by
\bes
\label{PB-spin}
\begin{align}
    \{G,E\}^{0}&= \{G^0,E^{0}\}\, , 
    \\
    \label{PB-spatial}
    \{G,E\}^{i}&= \{G^0,E^{i}\}+ \{G^i,E^0\}-\epsilon^{ijk}G_{j}E_k\, , 
\end{align}
\ees
One way to understand those generalized P.B.~\eqref{PB-spin} is to calculate the commutators of $\psi^\dagger(x)\s^{i}\psi(y)$ for a two-component Fermionic operator $\psi$, where $\s$ is the Pauli matrix, see ref.~\cite{Delacretaz:2022ocm} for details. 
Denoting $H=(\e,\bm\zeta), F=(f,\vg)$ where $\boldsymbol{\zeta}$ represents the spin-dependent counterpart of $\epsilon$, it is natural to extend eq.~\eqref{Boltzmann-f} as
\begin{align}
    \pd_{t}F +\{F,H\}=0\, . 
\end{align}
Explicitly, E.o.M. for $\{f,\vg\}$ that we shall employ reads
\bes
\label{eq-fg}
\begin{align}
\label{f-eq}
&\,\partial_{t}f  +\{f,\e\}=0\, ;\\
\label{s-eq}
&\,\partial_{t} \vg +\{\vg,\e\}+\{f,{\bm \zeta}\}- {\vg}\times {\bm \zeta}=0\, .
\end{align}
\ees
We assume that \(\vg\) has no background value, and consequently, neither does \(\vzeta\).
In the absence of $\vzeta$, the evolution of $\vg$ is determined by $\e$ in the same way as $f$. 
The presence of $\bm\zeta$ introduces the coupling between $\vg$ and $f$ in their evolution. 
The term $\boldsymbol{g} \times \boldsymbol{\zeta}$ originates from the last term in eq.~\eqref{PB-spatial} which is a manifestation of $SU(2)$ algebra of spin $1/2$ particle. 
This non-linear term implies that if $\boldsymbol{\zeta} \propto \boldsymbol{B}$, $\boldsymbol{g}$ will evolve in time unless it aligns with the magnetic field $\vB$ even if other terms like $\{\vg,\epsilon\}$ vanish.

Summarizing, we adopt a "bottom-up" viewpoint by proposing that dynamical variables of SKT, $(f,\vg)$, obey E.o.Ms which are expressible in terms of spin-dependent P.B.
Those equations have to be supplemented by the expression for the particle/spin energy $(\e, \vzeta)$. 
In general, $(\d\e, \vzeta)$ are functionals of $(f,\vg)$ and external fields, similar to the framework of Landau Fermi liquid theory.
The specification of those dependencies will be fixed by matching to the microscopic calculations; further details can be found in sec.~\ref{sec:matching}.

We conclude this subsection by discussing the gradient counting of those E.o.Ms. 
Because $f$ has non-zero background value while $\vg$ does not, 
it is consistent to count $(\d f,\d \e)\sim \sO(1), (\vg,\vzeta)\sim \sO(\pd)$.
Therefore, to determine the behavior of $\vg$ using eq.~\eqref{s-eq}, it is sufficient to input $\sO(1)$ contribution in $f$, which in turn is given by solving eq.~\eqref{f-eq}.
If one wishes to track the evolution of higher order contribution (e.g. $\sO(\pd)$ or $\sO(\pd^2)$) in $f$, additional contributions to eq.~\eqref{f-eq} are needed.
They include terms of the form $\{\vg,\vzeta\}$ (see Appendix~E of ref.~\cite{Delacretaz:2022ocm}), as well as modification of P.B. itself, since P.B. is the semiclassical limit of the Moyal bracket. 
We leave those refinements to future work (see e.g. ref.~\cite{Yang:2024sfp} on SKT with higher-gradient corrections).


\subsubsection{Schwinger-Keldysh action}
We now develop the action for spin kinetic theory in the Schwinger-Keldysh (SK) or closed-time-path formalism. 
In the near future, we wish to use the resulting SK action to describe both collisional effects and ensemble fluctuation.
Presently, we find that the SK formalism provides the most transparent description of the Noether current in phase space.

As reviewed in Refs.~\cite{Chou:1984es,Ghiglieri:2020dpq,Crossley:2015evo,Liu:2018kfw}, 
SK formalism introduces two copies of fields and expresses the action in terms of r-a variables.
Here, the r-variable will give the averaged behavior of observables, and the a-field describes quantum and stochastic effects. 
They are related to the average and the difference between the $1,2$-variables, respectively.
Those $1,2$-variables represent fields entering the forward and backward evolution in the path integral representation. 
The distribution functions appearing in E.o.Ms, etc, can be regarded as $r$-variables. 
For brevity, we retain only the $a$ subscript and omit the $r$ subscript when doing so does not cause confusion. 
SK action must obey the "unitarity" constraints~\cite{Crossley:2015evo}.
\begin{align}
\label{I-unitary}
I[\chi,\chi_a=0] &= 0,\quad
I[\chi,-\chi_{a}] =- I^{*}[\chi,\chi_a],\quad
\mathrm{Im}\, I \geq 0\,.
\end{align}
The first condition in eq~\eqref{I-unitary} implies that the effective action can be organized as an expansion in powers of the $a$-field $\chi$, starting at linear order in $\chi_a$.

We take $(f,\vg)$ as the r-variables in the action. 
As for a-variable, we use the set $\Phi_a=(\phi_a,{\bm\varphi}_a)$ rather than $(f_a,\vg_a)$. 
To linear order in a-variables and in the absence of dissipation (collisions),
the following reproduces E.o.M.~\eqref{eq-fg} upon varying $\Phi_a$ 
\begin{align}
\label{SK-action}
  I=\int_{t,\Gamma} \, &\phi_{a}
  \left[\partial_{t}f  +\{f,\e\}
\right]
\, \,+{\bm\varphi}_a\cdot
\left[
\partial_{t} \vg +\{\vg,\e\}+\{f,{\bm \zeta}\}- {\vg}\times {\bm \zeta}\right]
\end{align}
Note the quadratic and higher-order terms in a-field encode fluctuations (noise), but they are not important for the present discussion and hence will not be included.

While one might think that $\Phi_a$ merely acts as the Lagrangian multiplier that ensures only solutions to E.o.M. contribute to the path integral
\footnote{
If we were using Martin–Siggia–Rose (MSR) method~\cite{Martin:1973zz}, 
we would place $i\phi_{a}$ where $\phi_a$ appears. 
However, the second condition in \eqref{I-unitary} requires that we shall use $\Phi_{a}$ without the prefactor $i$.
},
we emphasize that both $\Phi_a$ and its r-variable counterpart $\Phi$ possess much richer physical content in this formulation.
This is reflected through the fact that various symmetry transformations can be implemented by the linear shift of $\Phi_{r/a}$.
Particularly, we are interested in $U(1)$ gauge transformation $A_{\mu}\to A_{\mu}+\pd_{\mu}\a(x)$ and space-time translation parametrized by $b_{\mu}$.
We further assume that $f$ describes the particles that carry the same $U(1)$ charge. 
By construction, $U(1)$ transformation is represented by (see also Ref~\cite{Delacretaz:2025ifh})
\bes
\label{U1-Tran}
\begin{align}
    \label{U1-phi}
\,\text{U(1)}:\, &   \phi_{r/a}(x)\to \phi_{r/a}(x)+\a_{r/a}(x)\, , 
    \qquad
    {\bm \varphi}_{r/a}\to {\bm \varphi}_{r/a}\, .
\\
\label{Tran-phi}
 \, \text{Translation}:\, & 
    \phi_{r/a}(x)\to \phi_{r/a}(x)+b_{r/a}(x)\cdot p\, , 
    \qquad
    {\bm \varphi}_{r/a}\to {\bm \varphi}_{r/a}\, . 
\end{align}
\ees
where the on-shell four-momentum is given by $p^{\mu}=(\e_{\vp},\vp)$ and where we resume the $r,a$ index. 
Also by construction, an infinitesimal change in $\Phi$ results in a change in $F=(f,\vg)$ through generalized P.B.
\begin{align}
\label{Phi-trans}
    \Phi \to \Phi +\delta \Phi\, 
    \qquad
    F\to F+ \{F,\d\Phi\}\, 
\end{align}
Or explicitly
\bes
\label{D-fg-CT}
\begin{align}
    \d f&=\{f,\d\phi\}\, , 
    \\
    \d \vg&=\{\vg,\d\phi\}+\{f,\d{\bm\varphi}\}-\vg\times \d {\bm \varphi}\, . 
\end{align}
\ees

The \( U(1) \) transformation of \( \phi_{r/a} \) in eq.~\eqref{U1-phi} reflects that these fields can be interpreted as the local phase of Fermionic particles. 
Regarding the translation, a homogeneous $\vb$ will generate a change $\d f= {\bm b}\cdot\vpd f$  due to eq~\eqref{D-fg-CT} and is in accordance with our expectation for the translation. 
Turning to the change induced by homogeneous  $b_0$, we apply \eqref{D-fg-CT} to find
\begin{align*}
    \d f= -b_0(\vv\cdot \vpd\, f)=b_0\pd_{t}f= f(t+b_0)-f(t)\, ,
\end{align*}
which justifies its interpretation as time translation, at least in a linear approximation. 
Those transformations~\eqref{U1-Tran} will be used to derive the phase space density of the vector current and the energy-momentum tensor in sec.~\ref{sec:constitutive}.

The action presented in ref.~\cite{Delacretaz:2022ocm} focuses on only one copy of the field, which corresponds to either \(\Phi_{1}\) or \(\Phi_{2}\). We direct the reader to ref.~\cite{Delacretaz:2022ocm} for a detailed explanation of the non-linear relationship between \(F = (f, \vg)\) and \(\Phi\). In our experience, expressing the action in terms of \(\Phi_a\) and \(F\) (rather than $\Phi_{r}$) is more convenient for practical applications, particularly for action up to linear order in the a-field.

\subsection{Constitutive relation for observables
\label{sec:constitutive}
}

\subsection{Set-up}

For the effective description to be useful, we should require that the observable of interest in real space $O(x)$ can be expressed in terms of the dynamical variables through the relation 
    \begin{align}
    \label{O-con}
   O(x)= \int_{\vp} \tilde{O}(f,\vg)\, (\Gamma)\, . 
\end{align}
Here $\tilde{O}$ is a functional of the dynamical variable $F=(f,\vg)$ and can involve derivative operators in the phase space that act on $F$. 
The temporal dependence of $\tilde{O}$ should not be explicit and is only through that of dynamical fields.
If $O(x)$ is some Noether current associated with the symmetry of the underlying theory, one can read $\tilde{O}$ from the EFT action. 
For general observables and in the absence of any microscopic details, we should write down the most general expression at the gradient order of interest, called the constitutive relation.
Note that any time derivative can be substituted for spatial derivatives using the equations of motion and hence will not be included.
In the rest of this paper, we will focus on observables at leading order in small gradients.

As discussed earlier, if we compute the response of $O$ to external disturbance, the non-analytic behavior in Fourier space should be fully matched with that from computing the behavior of $F$. 
Therefore, more precisely speaking, we only require \eqref{O-con} to match with the non-analytic/dynamical part of $O(x)$.

\subsubsection{Noether current in phase space
\label{sec:N-current}
}

In this section, we derive the phase space Noether current associated with the conservation law. 
The main idea is to consider the infinitesimal transformation of eq.~\eqref{U1-Tran} and then read the current in phase space; see also ref.~\cite{Delacretaz:2025ifh}. 
We are primarily interested in the $r$-current, which is conjugate to the change in the $a$-field. 
In this case, it is convenient to use 
\begin{align}
\label{f-current}
    \{a,b\}= \vpd\cdot(a\vpdp b)- \vpdp\cdot(a\vpd b)\, 
\end{align}
to express the Lagrangian density in the action~\eqref{SK-action} as
\begin{align}
\label{L-change}
    \sL= &\,\int_{\vp} \phi_{a}\,\left[\pd_{t}f+
    \vpd\cdot\left((\vpdp\e) f\right)
    +\vpdp\cdot\le(
    (-\vpd\e)f
    \ri)
    \right]
    \no \\
    +&
    \varphi_{a}^l\left[\pd_{t}g_l+
    \vpd\cdot\left((\vpdp\e) g_l+
    (\vpdp\zeta_{l})\,f\right)
    +\vpdp\cdot\le(
    (-\vpd\e)g_l+(-\vpd\zeta_l)f
    \ri)+\epsilon_{lij}g^i\zeta^j
    \right]
\end{align}
Terms in the brackets of $\vpdp(\ldots)$ in the second line of eq.~\eqref{L-change}, such as $(-\vpd\e)$, are naturally interpreted as the force.

We first determine the current associated with $U(1)$~\eqref{U1-phi}.
From eq.~\eqref{L-change}, we find that the change in the Lagrangian density takes the form
\begin{align}
    \sL = \int_{\vp} (\pd_{\mu}\sV^{\mu})\a \, ,
\end{align}
where we identify $\sV$ as the vector Wigner function since its integration over $\vp$ leads to the vector current in the real space. 
The constitutive relation for $\sV^{\mu}$ follows from the first line of eq.~\eqref{L-change}.
In the absence of the external field,  we obtain the explicit expression for the constitutive relation for the vector Wigner function
\begin{align}
\label{V-const}
    \sV^{\mu}_{\con}[f]= v^{\mu} f\, , 
\end{align}
where $\vv=\vpdp\epsilon_{\vp},\ v^{\mu}\equiv (1,\vv)$.
Note, the conservation of vector current in real space can be verified from the equation of motion for $f$~\eqref{kinetic-eq1}, which takes the form
\begin{align}
\label{div-sV}
    \pd_{\mu}\sV^{\mu}=\vpdp\cdot\le(
    (-\vpd\e)f
    \ri)\, . 
\end{align}

Next, we turn to the spacetime translation~\eqref{Tran-phi} where the variation of $\sL$ can be related to the density of the canonical energy-momentum tensor
\begin{align}
    \sL = \int_{\vp}(\pd_{\mu}\sT^{\mu,\nu})b_\nu \, .
\end{align}
Following similar steps, we find the expression
\begin{align}
\label{T-con}
    \sT^{\mu,\nu}_{\rm con}[f]=p^{\nu}v^{\mu}f\, . 
\end{align}
While those results in eqs.~\eqref{V-const} and \eqref{T-con} may seem familiar, we emphasize that they follow from identifying various symmetry transformations within the current action formulation.

\subsubsection{Constitutive relation for axial Wigner function}
Now, we demonstrate how to write down the constitutive relation for the axial Wigner function $\sA$.
This is a physical observable of primary interest since $\sA$ can be defined as the expectation value of field theory operators (see sec.~\ref{sec:matching}) and describes the phase-space distribution of spin. 
According to our counting scheme, $\sA\sim \sO(\pd)$, so it must be expressed in terms of the gradient of $f$ and $\vg$. 
The constitutive relation for other observables can be derived analogously.

Since the presence of the medium breaks the Lorentz invariance, we shall discuss $\sA^0$ and ${\bm \sA}$ separately, which are pseudo-scalar and pseudo-vector, respectively. 
We consider the most general expression for the constitutive relation that is consistent with rotation and spatial parity
\bes
\label{A-const-all}
\begin{align}
\label{A0-const}
    \sA^0_{\rm con}[\vg]&=c_0 \vv\cdot \vg\, .
\\
\label{A-const}
   {\bm \sA}_{\rm con}[\vg, f]&=c_1\vg_{\parallel}+c_2\vg_{\perp}+c_3\vv\times {\bm\pd}f \, ,
\end{align}
\ees
We have projected $\vg$ in the direction longitudinal and transverse to the three velocity $\vv$, i.e.,
$\vg_{\parallel}\equiv\hat{\vv}(\hat{\vv}\cdot \vg), \vg_{\perp}\equiv\vg-\vg_{\parallel}$. 
The coefficients $c_{0,1,2,3}$ are function of $\e_{\vp}$. 
We shall match eq.~\eqref{A-const-all} with the field theoretical calculations later.

\section{Linear response from kinetic theory
\label{sec:kin-lin}
}
We shall match the kinetic theory description of the medium's response to external disturbance with the field theory. 
In kinetic theory, the response is 
characterized by the deviations from the background distribution, i.e. $(\d f,\vg)$ where $f = f_0 + \delta f$
and $\vg$.
Likewise, the presence of external fields also induces the change in energy, i.e., $\d \epsilon$ in $\epsilon = \epsilon_{\vp} + \delta\epsilon$ and ${\bm \zeta}$. 
In our set-up, $\vg,\bm\zeta$ has no background value so we do not put $\d$ in those symbols for brevity.

To analyze linear response, we set up power counting to keep track of the external perturbation by introducing a small parameter $\d$. 
The background value of the distribution function $f_0, \sV^{\mu}_0\sim {\cal O}(\d^{0})$
while $(\d\e,\bm\zeta,\vg)\sim {\cal O}(\d)$. 
For consistency, the function that parametrizes the symmetry transformation is also ${\cal O}(\d)$. 
While both $\d$ and gradient are counted as small, they are independent. 
For example, among terms which are linear in $\d$, $\d f, \d \e\sim \sO(\pd^0)$ while $\vg,\bm\zeta\sim \sO(\pd)$.

To the lower non-trivial order in ${\cal O}(\d)$, the linearized kinetic equations~\eqref{eq-fg} take the form:
\begin{align}
\label{kinetic-eq1}
\partial_t \delta f + \{\delta f, \epsilon_{\vp}\} + \{f_0, \delta\epsilon\} &= 0\, , \\
\label{kinetic-eq2}
\partial_t \bm{g} + \{\bm{g}, \epsilon_{\vp}\} + \{f_0, \bm{\zeta}\} &= 0\, ,
\end{align}
In Fourier space, the solutions are given by:
\begin{subequations}
\label{sol-lin-kin}
\begin{align}
\label{sol-df}
\delta f(Q) &= \frac{\vv \cdot \vq}{v \cdot Q} \,f_{0}'\,\delta\epsilon(Q), \\
\label{sol-dg}
\bm{g}(Q) &= \frac{\vv \cdot \vq }{v \cdot Q} \, f_{0}'\,\bm{\zeta}(Q)\, ,
\end{align}    
\end{subequations}
where from now on we assume that $f_{0}$ is a function of of $\epsilon_{\vp}$ and $f'_{0}$ denotes the derivative with respect to $\epsilon_{\vp}$.
We note from eqs.~\eqref{sol-df},\eqref{sol-dg}  that $(\d f,g)$ are non-analytic function in frequency and wave-vector $\o,\vq$,
the point we shall return to in sec.~\ref{sec:matching}. 
Note also that $(\d f,\vg)$ in eq.~\eqref{sol-lin-kin} are proportional to $f'_0$, which in the low temperature limit only receives contributions from excitations in the vicinity of the Fermi surface. 
The general lesson here is that the present EFT describes hard particles that contribute most to the deviation from the background value.
The phase-space volume occupied by those excitations can be significantly smaller than that occupied by the particles in the bulk, indicating a reduction in the density of d.o.f.s governing the system's dynamics.

The solutions~\eqref{sol-df} and~\eqref{sol-dg}, combined with the constitutive relations for the vector and axial Wigner functions~\eqref{V-const} and~\eqref{A-const-all}, allow us to compute the system's response to external perturbation.

\section{Linear response from field theory
\label{sec:field}
}
In this section, we present the field theoretical calculations of the Wigner function of a generic Fermionic field $\hat{\psi}$ in response to the external gauge fields and metric fluctuations in the linearized regime.
Eqs.~\eqref{W-EM-0},\eqref{W-EM-1},\eqref{W-h-0},\eqref{W-h-1}, below, are main results of this section, 
Those will be used to calculate the induced vector and axial Wigner function.

\subsection{Set-up}

Following refs.~\cite{Heinz:1983nx,Mueller:2002gd}, we define the Wigner function in terms of the SK variables as
\begin{equation}
\label{W-def}
W(x,\sP)=-\int_{y} \,\sqrt{-g(x)}\, U(x,x_{+})\, \langle\hat{\psi}_{r}(x_{+})\hat{{\bar\psi}}_{r}(x_{-})\rangle\, U(x_{-},x)\,e^{-i \sP\cdot y}\, ,
\end{equation}
where $x_{\pm}=x\pm\frac{y}{2}$ with $x$ representing the middle point between $x_{1},x_{2}$ and $y$ being the difference.
The overall $-$ follows the convention.
The expectation value $\langle\ldots\rangle$ is formally determined by averaging over the density matrix which is not necessarily in equilibrium. 
The appropriate "gauge link" $U(x,x')$ has been implemented as (see e.g. ref.~\cite{Fonarev:1993ht})
\begin{align}
\label{U-L}
    U(x_{1},x_{2};A_{\mu},g_{\mu\nu})=\exp\left[-\int^{x_{1}}_{x_{2}}\, \left(iA_{\mu}(z)
    + \Gamma_{\mu}(z)\right)
    dz^{\mu}\right]
\end{align}
where the path of integration is customarily chosen as a straight line connecting $x_{1}$ and $x_{2}$. 
The term proportional to $A$ in eq.~\eqref{U-L} ensures the invariance of $W(x,\sP)$ under gauge transformation
\begin{align}
\label{U1}
    \psi(x)\to \tpsi(x)=e^{-i\a(x)}\psi(x)\, , 
    \qquad
    A_{\mu}(x)\to \tilde{A}_{\mu}(x)=A_{\mu}+\pd_{\mu}\alpha(x)\, . 
\end{align}
The spin connection $\Gamma_{\mu}$ is defined in terms of the tetrad $e^{\mu}_{b}$, which satisfies $g_{\mu\nu}e^{\mu}_{b}e^{\nu}_{c}=\eta_{bc}$:
\begin{align}
\label{spin-con-def}
    \Gamma_{\mu}=-\frac{i}{4}\sigma^{bc}
    \le(e^{\nu}_{b}\nabla_{\mu}e_{\nu,c}\ri)\, , 
\end{align}
and $\sigma^{bc}=\frac{i}{2}[\gamma^{b}, \gamma^{c}]_-$ are the generator of Lorentz transformation.
The spin-connection term ensures the covariance of $W(x,\sP)$ under local Lorentz transformations (LLTs).
The Wigner (Fourier) transform has been performed so that $W$
depends on both $x$ and the off-shell four-momentum $\sP_{\mu}=(p_{0},\vp)$ that is conjugate to $y$.
We included $\sqrt{-g(x)}$ so that the integral $\int_{y}$ is invariant under linearized diffeomorphism.
~\footnote{
Diffeomorphism covariance to non-linear order in the diffeomorphism transform parameter $\xi_{\nu}$ requires special care because the midpoint $x$ used in defining does not transform covariantly. While several covariant prescriptions exist, notably via introducing horizontal lift to $e^{-i \sP\cdot y}$~\cite{PhysRevD.37.2901,Fonarev:1993ht,Antonsen:1997dc, Liu2018,Hayata2020}, we have confirmed that these refinements yield identical results in our linear-response analysis and are therefore omitted for simplicity.
}

In the calculation below, we use the SK propagators $S_{ra}\sim \langle \psi_{r}\bar{\psi}_{a}\rangle, S_{rr}\sim \langle \psi_{r}\bar{\psi}_{r}\rangle$ etc of the unperturbed the system (see the appendix of ref.~\cite{Ghiglieri:2020dpq})
\begin{align}
\label{SK-propagator}
\begin{pmatrix}
S_{rr}(\sP) & S_{ra}(\sP) \\
S_{ar}(\sP) & 0
\end{pmatrix}
\,
= M(\sP)
\begin{pmatrix}
D(\sP) & \frac{i}{-\sP^{2}-m^{2}+i\epsilon p^{0}} \\
\frac{i}{-\sP^{2}-m^{2}-i\epsilon p^{0}} & 0
\end{pmatrix}\, . 
\end{align}
Here, we have purposely factored out the spin structure by introducing the matrix
\begin{align}
M(\sP)= -\slashed{\sP}+m\, . 
\end{align}
The medium's properties is encoded in $S_{rr}$ through 
\begin{align}
\label{DSs}
    D(\sP)&=\le(\frac{1}{2}-f_0(p^0)\ri)\,\delta_{\sP}\, , 
\end{align}
where $\d_{P}$ imposes the on-shell condition 
\begin{align}
\label{dp}
    \delta_{\sP}\equiv 2\pi \delta(\sP^{2}+m^2) 
\end{align}
and $f_0$ would be identified as the background distribution. 
We use the expression in the medium's rest frame so that the distribution function depends on $p^0$ alone. 
While in the equilibrium $f_{0}(p^0)$ becomes the Fermi-Dirac distribution, the current formulation allows $f_{0}$ to be an arbitrary homogeneous and static distribution,
 including those far from equilibrium.
Note, $S_{rr}$ coincides with unperturbed Wigner function upto to a overal "$-$" sign, i.e. $S_{rr}(\sP)=-W_{0}(\sP)$. 

\subsection{EM-response}

The linear response of Wigner function, $\delta W$, to an spacetime-dependent external gauge field $A_{\mu}$, in Fourier space takes the form
\begin{equation}
\label{delta-W}
\delta W(\sP,Q)=
G^{\mu}(\sP,Q)\, A_{\mu}(Q)
+\delta_{L}W(\sP,Q)
\end{equation}
The first term in eq.~\eqref{delta-W} is proportional to the retarded correlator 
\begin{align}
\label{G-EM}
    G^{\mu} \sim i\langle \psi_{r}\bar{\psi}_{r}\, J^{\mu}_{a} \rangle
\end{align}
where $J^{\mu}_a=J^{\mu}_1-J^{\mu}_2=\bar{\psi}_r \gamma^{\mu} \psi_a + \bar{\psi}_a \gamma^{\mu} \psi_r$ for $J^\mu = \bar{\psi}\gamma^\mu \psi$.
The gauge link~\eqref{U-L} gives rise to the second term in \eqref{delta-W}, which at linear response order reads
\begin{align}
\label{dW-L0}
    \delta_{L}\,W(\sP,Q)= -A\cdot\pd_{\sP}W_0(\sP)=A\cdot\pd_{\sP}S_{rr}(\sP)
\end{align}
where $\pd_{\sP}$ denotes the derivative with respect to $\sP$. 
Under gauge transformation~\eqref{U1}, 
eq.~\eqref{dW-L0} transforms as
$-(\pd \a)\cdot\pd_{\sP}W_0(\sP)$, 
which precise cancels the difference in phase of $\psi(x_1), \bar{\psi}(x_2)$ in the definition of the Wigner function~\eqref{W-def} to the first order in gradient.

We present results to the first non-trivial order in the perturbative calculation and up to the first order in the gradient expansion.
The correlator~\eqref{G-EM} is then given by:
\begin{equation}
\label{EM-loop}
G^{\mu}(\sP,Q)=i\,\left[S_{rr}(\sP_{+})\gamma^{\mu}\,S_{ar}(\sP_{-})+S_{ra}(\sP_{+})\gamma^{\mu}S_{rr}(\sP_{-})\right],
\end{equation}
where 
$Q_{\mu}=(\o,\vq)$ and $\sP_{\pm}=\sP\pm\frac{1}{2} Q$.
Substituting the explicit expression for the propagators~\eqref{SK-propagator} into eq.~\eqref{EM-loop} yields
\begin{align}
\label{G-I}
G^{\mu}&=-M(\sP_{+})\g^{\mu}M(\sP_{-})\,
\le[ 
\frac{D(\sP_{+})}{-\sP^{2}_{-}-m^{2}-i\epsilon p^{0}_-}
+
\frac{D(\sP_{-})}{-\sP^{2}_{+}-m^{2}+i\epsilon p^{0}_{+}}
\ri]
\no \\
&=
-M(\sP_{+})\g^{\mu}M(\sP_{-})\,
\left[
\frac{D(\sP_{+})-D(\sP_{-})}{
2\sP\cdot Q-i\epsilon p^{0}}
\right]
 \no \\
    &=  -\left[\left(2\sP^{\mu} M(\sP) +(\sP^{2}+m^{2})\,\g^{\mu}\right)+\frac{-i}{2} [M(\sP),\sigma^{\mu\a}Q_{\a}]_+ \right]
\frac{Q\cdot \pd_{\sP}D(\sP)}{2\sP\cdot Q-i \epsilon p^{0}}
\end{align}
The simplification from the first to the second line employs the relation
\begin{align}
&\,   \frac{i\,  \delta_{\sP_{+}}}{-\sP^{2}_{-}-m^2-i \epsilon p^{0}_{-}}
    =   \frac{i\,\delta_{\sP_{+}}}{-\sP^{2}_{+}-m^2+(\sP^{2}_{+}-\sP^{2}_{-})-i \epsilon p^{0}_{-}}
    =  \frac{i\,\delta_{\sP_{+}}}{2 \sP\cdot Q-i \epsilon p^{0}_{-}}\, . 
\end{align}
From the second line to the third line, we expand in $\o/|p_{0}|, q/|p_{0}|$ as $p^0\sim T_{\rm eff}$ (c.f. eq.~\eqref{eq:domain})  up to first order in this gradient expansion.
Turning to $\d_L W$, we have
\begin{align}
\label{L-A}
    \d_{L}W= A_{\mu}\cdot\left[M(\sP)\pd_{\sP}^{\mu}D(\sP)-\g^{\mu}D(\sP) \right]\, , 
\end{align}
which only contributes at zeroth order in the gradient expansion.

We now collect eqs.~\eqref{G-I}, \eqref{L-A} to obtain the full expression for the response.
Keeping terms up to zeroth order in eq.~\eqref{G-I} yields:
\begin{align}
\label{GEM-0}
   G^{\mu}_{0}= -M(\sP)\sP^{\mu}\,\left( \frac{Q\cdot \pd_{\sP}D(\sP)\,}{\sP\cdot Q }\right)+\g^{\mu} D(\sP)
\end{align}
where the second term originates from the $(\sP^2 + m^2)\gamma^{\mu}$ contribution in eq.~\eqref{G-I}. 
Here and hereafter, we will omit $i \epsilon p^{0}$ for brevity. 
Adding $\d_{L}W$~\eqref{L-A} gives the expression
\begin{align}
\label{W-EM-0}
    (\d W)_0 &= -M(\sP) \le[\frac{\sP^{\mu}(Q\cdot\pd_{\sP}D(\sP))}{\sP\cdot Q}
    -
    \pd_{\sP}^{\mu}D(\sP)
    \ri]\, A_{\mu}
    \no\\
    &=M(\sP)
    \left[
    \sP^{\mu}\,\frac{Q\cdot\pd_{\sP}f_0}{\sP\cdot Q}
    -\pd_{\sP}^{\mu}\,f_0
    \right]\d_{\sP}\, \times A_{\mu}\, . 
\end{align}
The derivative $\partial_{\sP}D(\sP)$ in the first line contains contributions from both $f_0$ and $\delta_{\sP}$, but the terms involving derivatives of $\delta_{\sP}$ cancel exactly, yielding the second line. 
This cancellation indicates that the system's response to the gauge field $A_\mu$ should depend solely on how the particles' distribution is modified ($\propto \pd_{\sP}f_0$), not on variations of the mass-shell condition $\propto \pd_{\sP}\d_{\sP}$ at this order. 
Additionally,
we observe a precise cancellation between the terms proportional to $\gamma^{\mu} D(\sP)$ appearing in both $G^{\mu}_0$ and $\delta_L W$.
In $G^{\mu}$, this term originates from the coupling of $A_{\mu}$ to the $\gamma^{\mu}$,
while in $\delta_L W$ it arises from the $\slashed{\sP}$ structure of $W_0$ when computing the gauge link variation. 
Since the gauge invariant $W$ is a function of the combination $(\sP+A)$, they cancel.

The first-order result only receives the contribution from the $G^{\mu}$ and is expressed as follows:
\begin{align}
\label{W-EM-1}
   (\delta W)_1
    &= 
     -\frac{1}{8}
    [
    M(\sP),\sigma^{\mu\nu}F_{\mu\nu}
    ]_+\, \times
    \frac{Q\cdot \pd_{\sP}D(\sP)}{\sP\cdot Q}
    \, . 
\end{align}
Observe that $(\delta W)_1$ depends on field strength and hence is manifestly gauge invariant.

\subsection{Gravity response}

We now extend our analysis to a metric perturbation around the Minkowski metric $\eta_{\mu\nu}$ (see also Refs.~\cite{Hayata2020,Lin:2024svh})
\begin{align}
\label{metric-per}
g_{\mu\nu}=\eta_{\mu\nu}+h_{\mu\nu}\, ,
\end{align}
In parallel to the analysis of the response to EM fields, the change of $W$ reads
\begin{equation}
\label{deltaW-h}
\delta W(\sP,Q)=
G^{\mu\nu}(\sP,Q)\,h_{\mu\nu}(Q)
+\delta_{L}W(\sP,Q)\, . 
\end{equation}
where the correlator is defined by
\begin{align}
\label{GR-h}
    G^{\mu\nu}\sim -\frac{i}{2}\langle \psi_{r}\bar{\psi}_{r}\ \hat{T}^{\mu\nu}_{a, {\cal T}} \rangle\, . 
\end{align}
Here symmetrized (Belinfante) EMT is explicitly given by
\begin{align}
\label{Ta}
T^{\mu\nu}_{a}
&=\frac{i}{2}\le[\bar{\psi}_{r}\g^{(\mu}\overleftrightarrow{\nabla}^{\nu)}\psi_{a}+\eta^{\mu\nu}\bar{\psi}_{r}
(\g\cdot\overleftrightarrow{\nabla}-m)\, \psi_{a}
\ri]+(r\leftrightarrow a)\, .
\end{align}
where $A^{(\mu}B^{\nu)} \equiv \tfrac{1}{2}(A^\mu B^\nu + A^\nu B^\mu)$ denotes symmetrization.
In EMT~\eqref{Ta}, the part proportional to $\eta^{\mu\nu}$ (times Lagrangian density for Dirac fermions) is refered to as the scalar part of EMT and the remaining tensor part is denoted by $T^{\mu\nu}_{a,{\sT}}$ which is used in the correlator eq.~\eqref{GR-h}.
One can check that the contribution of the scalar part to the response represents the change in the integration measure in the presence of the metric perturbation and therefore would be exactly canceled by the variation of $\sqrt{-g}$ in the construction of $W$~\eqref{W-def}. 
We have also verified this cancellation through explicit calculation. 
Therefore, in eq.~\eqref{deltaW-h}, we retain only the tensor part $T^{\mu\nu}_{a,\mathcal{T}}$.

Following steps similar to calculating $G^{\mu}$, we obtain the expression 
\begin{align}
\label{G-T}
G^{\mu\nu}(\sP,Q)&= -\frac12\,\sP^{(\nu}G^{\mu)}(\sP,Q),
\end{align}
where $G^\mu$ is given by eq.~\eqref{G-EM}.
For the gauge link contribution $\delta_L W$, we need to compute the variation of the tetrad field under metric perturbations, $\delta e^{\mu}_{b} = -\tfrac{1}{2} h^{\mu}_{\ \nu} e^{\nu}_{b}$, yielding:
\footnote{
The variation of the tetrad $\delta e^{\mu}_{b}$ is not uniquely determined by $h_{\mu\nu}$ alone, due to the freedom of local Lorentz transformations on $e$ that leave the metric invariant. This ambiguity is resolved when computing physical observables, such as the vector Wigner function, by properly accounting for the tetrad dependence of the gamma matrices. 
See the master thesis.~\cite{mo2026bottom} by one of us (Z.L.M) for details.
}
\begin{align}
\label{L-h}
    \delta_{L}W = -\frac{i}{8} \left[ \partial_{\sP}^{\mu} W_{0}(\sP), \sigma^{\nu\rho} Q_{\rho} \right]_+ h_{\mu\nu}+\sO(\pd^2).
\end{align} 
In contrast to the case of EM perturbation~\eqref{L-A}, the gauge link contribution is first order in gradient.

With eqs.~\eqref{G-I} and \eqref{L-h} at hand, we obtain the response of $W$ to the metric perturbation at ${\cal O}(\pd^0)$ and ${\cal O}(\pd^0)$
\begin{align}
\label{W-h-0}
    (\d W)_{0}
    &=\frac12(M(\sP)\sP^{\mu}\sP^{\nu}h_{\mu\nu})\frac{Q\cdot\pd_{\sP}D}{\sP\cdot Q}
    -\frac12D(\sP) \sP^{\mu}\g^{\nu}h_{\mu\nu}\, .
\\
\label{W-h-1}
    (\d W)_1 &=-\frac{i}{8}[M(\sP),\sigma^{\mu\a}Q_{\a}]_+\,\left[\sP^{\nu}\frac{Q\cdot\pd_{\sP}D(\sP)}{\sP\cdot Q}-\pd_{\sP}^{\nu}D\right] h_{\mu\nu}\, . 
\end{align}
Note $[\ldots]$ terms in $(\d W)_1$ is transverse to $Q_{\nu}$ and $\sigma^{\mu\a}Q_{\a}$ is transverse to $Q_{\mu}$.
Therefore, $(\d W)_1$ is manifestly invariant under the linearized diffeomorphism parametrized by $\xi_{\mu}$, i.e., $h_{\mu\nu}\to h_{\mu\nu}+2\pd_{(\mu}\xi_{\nu)}$. 
This is expected due to the inclusion of the gauge link contribution.

\section{Matching
\label{sec:matching}
}
\subsection{Set-up}
The main purpose of this work is to demonstrate that SKT can be formulated using the "bottom-up" approach as outlined in sec.~\ref{sec:eff}. 
To do that, we have formulated the EFT for spin kinetic theory in sec.~\ref {sec:spin-action} and obtained the constitutive relation for vector and axial Wigner functions in sec.~\ref {sec:constitutive}. 
Those constitutive relations together with the linearized solutions to SKT in Sec~\ref{sec:kin-lin} describe linear response from the EFT side. 
In this section, we show that SKT results can be matched to a field theoretical analysis of Wigner function~\eqref{W-def} in sec.~\ref{sec:field}. 
This serves as a non-trivial test of our construction.

We shall concentrate on the vector and axial Wigner function for particles, which can be determined from the field theory side through the relation
\begin{align}
\label{V-A-def}
&\,    \sV^{\mu}(x,\vp)=-\frac12\int_{p^0}\, {\rm \tr}
    \le[ \g^{\mu}W(x,\sP)\ri]\theta(p^0)
    ,\qquad
    \sA^{\mu}(x,\vp)=-\frac12\int_{p^0}\, {\rm \tr}
    \le[ \g^{\mu}\g^{5}W(x,\sP)\ri]\theta(p^0)\, , 
\end{align}
where the prefactor $1/2$ averages over the spin state and $\theta(p^0)$ is included in the integration to pick-up particle contribution~\footnote{Apparently, $\sV^{\mu},\sA^{\mu}$ are not Lorentz (pseduo)vector (recall $f$ is not Lorentz scalar), but $\e_{\vp} \sV^{\mu}$ etc are.}. 
Note $\sV^{\mu},\sA^{\mu}$ here are function of three-momentum $\vp$, 
in accordance with our definition for distribution functions $(f,\vg)$. 
Before moving on, let us check the background expression obtained from eq.~\eqref{V-A-def} using $W_0=-S_{rr}$
\begin{align}
\label{VA-0}
    \sV^{\mu}_0 = v^{\mu}\, f_0\, ,
    \qquad
    \sA^{\mu}_0 =0
\end{align}
Here and hereafter, we dropped $1/2$ from the combination $(1/2-f_0)$ (see eq.~\eqref{DSs}) and retained only the $f_0$ term, which describes the medium's effect. 
The above result~\eqref{VA-0} is consistent with our set-up that the background particle distribution and the spin-dependent distribution are $f_0$ and ${\bm 0}$, respectively.

We are interested in showing that the non-analytic part of the $(\d \sV, \sA)$ is expressible in terms of the distribution function $(\d f, g)$ obtained in SKT. 
For this to happen, we first need to verify both $(\d \sV,\sA)$ and $(\d f, g)$ share the common non-analytic structure. 
Indeed, the former takes form (see eqs.~\eqref{W-EM-0},\eqref{W-EM-1},\eqref{W-h-0},\eqref{W-h-1})
\begin{align}
\label{W-non}
   \d W_{\rm n.a.}\sim \frac{Q\cdot\pd_{\sP}f_0}{\sP\cdot Q}=\frac{(Q\cdot u) f'_0}{\sP\cdot Q}\, ,
\end{align}
where $u^{\mu}=(1,0,0,0)$ is the four velocity of the medium at rest.
Meanwhile, 
the linearized solution for SKT~\eqref{sol-lin-kin} can be collectively written as
\begin{align}
\label{fg-non-ana}
        (\d f, \vg)=\frac{\vv\cdot\vq }{v\cdot Q}\,f'_0\times(\d \e,\vzeta)\, , 
\end{align}
Upon integration over $\int_{p^{0}}$, $\d W_{\rm n.a.}$~\eqref{W-non} leads to the contribution that is proportional to $1/(v\cdot Q)$, as in the kinetic theory results~\eqref{fg-non-ana}.

We shall elaborate that the identification of the non-analytic part of a given observable is inherently ambiguous. 
This ambiguity arises from the freedom to add or remove an analytic component without altering the non-analytic structure. 
For example, the relation
\begin{align}
\label{n-sep}
    \frac{Q\cdot \pd_{\sP}f_0}{v\cdot Q} =
    -\frac{\vv\cdot\vq}{v\cdot Q}f'_0+ f'_0\, .
\end{align}
indicates that both its L.H.S. and the first term can be referred to as the non-analytic part. 
Nevertheless, the physical observables are given by the sum of the analytic and non-analytic pieces and have no ambiguity at all.

In fact, we can employ the aforementioned freedom to identify the non-analytic component, thereby making the matching more transparent. 
The numerator in the expression for $(\d f,\vg)$ is proportional to the spatial gradient because of the P.B. structure in the kinetic equation. 
On the other hand, the numerator of $(\d \sV,\sA)_{\rm n.a.}$~\eqref{W-non} is proportional to $(Q\cdot u)f'_0$, originating from a change in $p^0$ by the amount $\o$ due to the temporal variation of external fields.
Despite this difference, both expressions share the same non-analytic behavior upon using eq.~\eqref{n-sep} (i.e., upon integrating over $\int_{p^0}$).

We shall illustrate the matching for response $(\d \sV, \sA)$ induced by both gauge field and metric perturbation.
While the explicit forms of $(\delta\epsilon, {\bm \zeta})$ depend on the specifics of the sources, 
the constitutive relations and their transport coefficients are expected to be universal. 
This reflects the fact that the relation between observables and dynamical variables does not depend on how the profile of the latter is created.
An explicit verification of this anticipation, hence, provides a non-trivial check of our bottom-up formulation of SKT.

\subsection{Vector Wigner function
\label{sec:V-match}
}
The induced vector Wigner function from the gauge field perturbation can be obtained by substituting eq.~\eqref{W-EM-0} into the definition~\eqref{V-A-def}:  
\begin{align}
\label{V-field}
    \delta\sV^{\mu}_{\EM}&= -2\,\int_{p^0}
\sP^{\mu}A^{\rho}
\left[
\sP_\rho\frac{Q\cdot \pd_{\sP}f_0}{\sP\cdot Q}
- u_{\rho} f'_0(p^0)
\right]\d^{+}_{\sP}\,
\no \\
&=\int_{p^0}
v^{\mu}\,A^{\rho}
\left[
v_\rho\frac{\vv\cdot \vq}{v\cdot Q}\,f'_0(\e_{\vp})-(v-u)_\rho f'_0(\e_{\vp})
\right](2\pi)\d(p^0-\e_{\vp})
\no \\
&=
v^{\mu}\frac{\vv\cdot \vq}{v\cdot Q}f'_0(\e_{\vp})\,(v\cdot A)-  v^{\mu}f'_0(\e_{\vp})\,(\vv\cdot\bm{A})
\end{align}
where $\d^{+}_{\sP}=(\pi/\e_{\vp}) \d(p^0-\e_{\vp})$ denotes particle part of $\d_{\sP}$.
From the first line to the second, we have used eq.~\eqref{n-sep}.
In parallel, the result due to metric perturbation can be obtained from eq.~\eqref{W-h-0}:
\begin{align}
\label{V-metric}
\d\sV^{\mu}_{\rm G} =&\,  \int_{p^0}
(-h_{\alpha\beta}\sP^{\alpha}\sP^{\beta})
v^{\mu}
\frac{\vv\cdot \vq}{ v\cdot Q}\,f_0^\prime(p^0)\d^{+}_{\sP}
\no \\
&+
\left[(h_{\alpha\beta}\sP^{\alpha}\sP^{\beta})\left(
 v^{\mu} f'_0(p^0)-f(p^0) \pd_{\sP}^{\mu}\right)
-h^{\mu\nu}\sP_{\nu}\,f_0^\prime(p^0))
\right]\,\d^{+}_{\sP} \, ,
\no \\
=&\
 v^{\mu} \frac{\vv\cdot \vq}{v\cdot Q} f_0'\,(-\frac12\,\e_{\vp}\,h_{\alpha\beta}v^{\alpha}v^{\beta})+\textrm{analytic piece}
\end{align}
The first line is non-analytic, while the second line is analytic.
For the later, the first and second term represent the modification $\d p^0\propto \sP_{\mu}\sP_{\nu}\,h^{\mu\nu}$ that in turn changes $f_0(p^0)$ and $\d_{\sP}$ while the third term originates from the variation of the gamma matrices in curved spacetime.
After performing the integration, we obtain the third, where we only keep the non-analytic part of the contribution explicitly.

The matching~\eqref{V-field} with kinetic theory is now straightforward by noting eq.~\eqref{fg-non-ana}.
For EM response, we find
\begin{align}
\label{V-const-EM}
    \d \sV^{\mu}_{\EM}= \sV^{\mu}_{\con}[\d f_{\EM}] - v^{\mu}(\vv\cdot\vA)f'_0\, , 
    \qquad
    \d \e_{\EM}  = v\cdot A\, . 
\end{align}
where the constitutive relation $\sV_{\con}$ is given by eq.~\eqref{V-const} and where $\d f_{\EM}$ is determined by solving the kinetic equation in the presence of induced energy $\d \e_{\EM}$. 
This equivalence between kinetic theory for spin-averaged distributions and linear response theory for $\sV^{\mu}$ under gauge field perturbation has been established in various many-particle systems~\cite{Son:2012zy,Chen2016,Caron-Huot:2007cma,Gao:2020ksg,Delacretaz:2025ifh}.

Turning to the response to the metric perturbation~\eqref{V-metric}, we observe that the matching works equally well 
\begin{align}
\label{V-const-G}
    \d \sV^{\mu}_{\rm G}= \sV^{\mu}_{\con}[\d f_{\rm G}] +\textrm{analytic piece} \, , 
    \qquad
    \d \e_{\rm G}  = -\frac12\,\e_{\vp}\, h_{\mu\nu}v^\mu v^\nu
\end{align}
We note by using $g^{\mu\nu}p_\mu p_\nu = -m^2,g^{\mu\nu}=\eta^{\mu\nu}-h^{\mu\nu} $ that the energy change due to metric variations is precisely given by $\d\e_{\rm G}$ in eq.~\eqref{V-const-G}.
Remarkably, $\d f_{\rm G}$ is different from $\d f_{\EM}$, but the constitutive relation remains the same.

We investigate how the form of $\d \e$ can be determined by considering the transformation of $\d f$.
First, we notice that under $U(1)$~\eqref{U1-phi}, $\d f$ transforms as
\begin{align}
\label{df-U1-lin}
    \d f\to \d f +\{f_0(\e_{\vp}),\a(x)\}+\sO(\d^2)= \d f -(\vv\cdot\vpd\a)\,f'_0\, , 
\end{align}
On the other hand, $\d f_{\rm EM}\propto \d \e_{\EM}$~\eqref{sol-df} can be viewed a function of $A$ through the dependence of $\d \e_{\EM}$ on $A$.
Therefore, the change of $\d f_{\EM}$ that is induced by the change of $A$ under $U(1)$ should be matched with eq.~\eqref{df-U1-lin}.
For this to be true, we should require that under $U(1)$
$\d \e_{\EM}\to \d\e_{\EM} +i(\vv\cdot Q) \a$ which  uniquely fixes the expression of $\d \e$ and agrees that in eq.~\eqref{V-const-EM}.

For the metric perturbation case, we notice that the gradient of $\d \e_{G}$ in eq.~\eqref{V-const-G} is precisely the gravitational "force" obtained from the geodesic equation of an individual particle, i.e., $-\partial_i \epsilon_{\rm G} = \frac12 \e_{\vp}\,\partial_{i}(h_{\mu\nu}v^{\mu}v^{\nu}) = \e_{\vp}\Gamma^{\rho}_{i\mu}v^{\mu}v_{\rho}$ where $\G^{\rho}_{\mu\nu}$ is the Christoffel symbol.
The linearized kinetic equation can be written as
\begin{align}
    (v\cdot \pd)\d f+ \e_{\vp}v^{\mu}v_{\r}\Gamma^{\rho}_{i\nu}v^{i}f'_0(\e_{\vp})=0\, 
\end{align}
upon using $\partial^{\nu}_{\sP}f_0(\e_{\vp})=(0,\vv)\,\times f'_0(\e_{\vp})$. 
Define the standard covariant derivative under diffeomorphism that acts on the function in the phase space as~\footnote{
Since \( p_{\mu} \) transforms as a one-form under diffeomorphism, it induces a change in the function \( f \) so that $f$ does not transform as a scalar. This change is compensated by the change in the Christoffel symbols in the definition of eq.~\eqref{D-cov}.
}
\begin{align}
\label{D-cov}
    D_{\mu}\equiv \partial_{\mu} +p_{\rho}\Gamma^{\rho}_{\mu\nu}\,\partial^{\nu}_{\sP}
    \, , 
\end{align}
the kinetic equation can be recast as
\begin{align}
    v\cdot D f=0
\end{align}
Reversing the logic, this implies that diffeomorphism covariance also fixes the form of $\d\e_{\rm G}$.

We further note that local symmetries, such as $U(1)$ and diffeomorphism, are also sufficient to determine the analytic piece of the expression for $\d \sV^{\mu}$ at zeroth order in the gradient expansion for both EM and metric perturbations.
Let us take $\d \sV^{\mu}_{\EM}$ as an exmaple. 
First, we observe that the analytic piece, if non-zero, must be gauge-dependent, since the gauge-invariant $F_{\mu\nu}$ is already $\sO(\pd)$.
Applying the transformation of $\d f$~\eqref{df-U1-lin} to the expression $V_{\EM}$~\eqref{V-const-EM}, we have already seen the form of the analytic piece (the second term) is fully fixed to cancel the change in $\sV_{\con}$ due to $\d f$. 
We have also examined how the diffeomorphism covariance implies the presence of an analytic piece (i.e., the second line of eq.~\eqref{V-const-G}).

We end this subsection by asking what if we were not using $\d f$, which is not gauge invariant, but choose a gauge invariant distribution, defined as,
\begin{align}
\label{f-cov}
    \d f_{\rm cov}\equiv \d f-\vv\cdot\vA\,f'_0\, , 
\end{align}
as our dynamical d.o.f.
Then the expression for $\sV^{\mu}$ simplifies
\begin{align}
    \d\sV^\mu_{\rm EM}=\sV_{\con}^\mu[\d f_{\rm cov}]\, . 
\end{align}
As a price, the linearized equation for $f_{\rm cov}$ no longer preserves the Poisson bracket structure~\eqref{f-eq}. It instead takes the form
\begin{align}
\label{fcov-eq}
    (v\cdot \pd) f_{\rm cov}+ (\vv\cdot \vE)f'_0=0
\end{align}
which gives the solution
\begin{align}
\label{df-cov}
    \d  f_{\rm cov}= -i\frac{\vv\cdot\vE}{v\cdot Q}\,f'_0\, . 
\end{align}

The lesson we learned here is that the effective description of kinetic theory enjoys a certain flexibility in choosing the dynamical variable. 
If we wish to simplify the structure of E.o.M, we use $f$ defined through eq.~\eqref{f-eq}, which is expressible in terms of P.B. 
Alternatively,
to highlight gauge invaraince, we take $f_{\rm cov}$.
Both choices are equally acceptable as they differ by an analytical piece.
The choice between them depends on the specific context.

We close with an observation that we find interesting and hope it could prove useful in broader contexts. 
While the conventional wisdom insists on imposing gauge invariance when constructing EFT actions, 
the action we employ~\eqref{SK-action} is not gauge invariant. 
However, this does not pose a problem in this instance because the gauge transformations introduce only an analytic piece in \(V_{\con} \), as we have demonstrated above. 
This suggests that the requirement for gauge invariance may be overly restrictive in certain scenarios. 
We may leverage this flexibility to simplify the analysis by considering gauge-dependent dynamical variables.

\subsection{Axial Wigner function}
One of the objectives in introducing $\vg$ is to parametrize the axial Wigner function $\sA$, which characterizes the spin distribution arising from inhomogeneities and perturbations. 
In this subsection, we demonstrate how this is achieved for $\sA$ computed from field theory. 
At the first non-trivial gradient order, the constitutive relation for $\d \sV$ only depends on $f$. 
In contrast, $\sA$ can depend on both $\vg$ and $\vpd f$, as detailed in expression~\eqref{A-const}. 
Thus, exploring this sector is both non-trivial and rich in physics.

We first evaluate the response to the EM perturbation $\sA_{\EM}$ and metric perturbation $\sA_{\rm G}$. 
Using eq.~\eqref{V-A-def} and eq.~\eqref{W-EM-1}, we find
\begin{align}
\label{A-field-int}
 \sA^\mu_{\EM}  &= 2\int_{p^0}
\tF^{\mu\nu}
\left[-v_{\nu}
\frac{\vv\cdot \vq}{ v\cdot Q}f_0^\prime(p^0)
+ v_{\nu} f'_0(p^0)-f(p^0) \pd^{\sP}_{\nu}\right]\d^{+}_{\sP}
\no\\
 &=\frac{\vv\cdot \vq}{v\cdot Q}\,f'_0\,(-\frac{\tF^{\mu\nu}v_{\nu}}{\e_{\vp}})\, 
    + \frac{f_0}{\e_{\vp}^2}\, \tF^{\mu\nu}\,v^{\perp}_{\nu}\, . 
\end{align}
where $v_\perp\equiv v-u$.
We have used the identity
\begin{align}
\label{M-rotation}
 [M(\sP),\sigma^{\nu\a}Q_{\a}]_{+}=2\le(
    \e^{\nu\sigma\a\b}\g^{5}\g_{\sigma}\sP_{\a}Q_{\beta} + m \sigma^{\nu\a}Q_{\a}
    \ri)\, 
\end{align}
when taking the trace.
Turning to $\sA_{G}$, we have
\begin{align}
\label{A-metric}
\sA^{\mu}_{\rm G} = 
\frac{\vv\cdot\vq}{v\cdot Q}\,f_0'\,(\frac i2\epsilon^{\mu\sigma\alpha\beta} v_{\alpha}Q_{\beta}\,h_{\sigma\nu}v^{\nu})
-f'_0\,  (\frac i2\epsilon^{\mu\sigma\alpha\beta} v_{\alpha}Q_{\beta}h_{\sigma\nu} (v^{\nu}-u^{\mu})) \, \, .
\end{align}
We can derive this expression by substituting $\sP^{\mu}A^{\rho}$ in the $\d \sV_{\EM}$~\eqref{V-field} with $\epsilon^{\mu\sigma\alpha\beta}\sP_{\alpha}Q_{\beta}h_{\sigma}^{\, \rho}$ (multiplied by an appropriate overall constant), inspired by the similarities between eqs.~\eqref{W-EM-0} and ~\eqref{W-h-1}.

We then find via explicit calculation that those expressions for $\sA$ from the field theory calculations can be matched to the kinetic theory description as follows
\bes
\label{AEM-match}
\begin{align}
\label{AEM-relation}
{\bm \sA}_{\EM}&=
{\bm \sA}_{\con}[\vg_{\EM}, f_{\rm cov}]
- f'_0\,\frac{\vv^2}{\e_{\vp}} \vB_{\perp} - \frac{f_0}{\e_{\vp}^2} (\vv\times \vE) \, ,
\\
\label{AG-relation}
\bm\sA_{\rm G} &= {\bm \sA}_{\con}[\vg_{\rm G},f_{\rm G}]+\left[- \vv^2 \vB^{h}_{\perp} + (\vv\times \vE^h) 
+  \vv\times\nabla(\vv\cdot\vA^h)-\vB^{h}
\right]\, f'_0 + \Delta {\bm\sA}_{\rm G}\, , 
\\ 
\sA^{0}_{\EM}&=
\sA^0_{\rm con}[\vg_{\EM}]+\frac{f_0}{\e^{2}_{\vp}}(\vv\cdot\vB)\, , 
\\
\sA^{0}_{\rm G}&=\sA^0_{\rm con}[\vg_{\rm G}]-f_0^\prime(\vv\cdot\vB^h)+\Delta \sA^{0}_{\rm G}\, , 
 \end{align}
 \ees
 where $\vg$ are obtained from eq.~\eqref{fg-non-ana} with the spin energy given by
 \bes
 \label{zeta-loop}
\begin{align}
 \label{zeta-EM}
{\bm\zeta}_{\EM} &= -\frac{\vB}{\e_{\vp}}\, . 
\\
\label{zeta-G}
{\bm \zeta}_{\rm G} &= \vB^{h}\, .
\end{align}
\ees
Here, we have defined the analog gauge fields induced by metric perturbations (but with a different mass dimension):
\begin{align}
A^{h}_{\mu} = \frac12 h_{\mu\rho}v^{\rho}, \quad 
\vB^{h} = \nabla\times \vA^{h}, \quad 
\vE^{h} = \nabla A^h_0 - \partial_{t}\vA^{h}\, .
\end{align}
We also defined $\Delta \sA^{\mu}_{\rm G}\equiv f'_0(\frac i2\epsilon^{\mu\sigma\alpha\beta} v_{\alpha}Q_{\beta}h_{\sigma 0})$, coming from the term proportional to $u^{\mu}$ in eq.~\eqref{A-metric}.
Importantly, the specific values of coefficients ($c$s) in $\sA_{\rm con}$~\eqref{A-const} do NOT depend on the types of perturbations and are universally given by
\begin{align}
    \label{cs}
    c_0=c_1=1\, , 
\qquad
c_2=(1-\vv^2)\, ,
   \qquad
c_3=\frac{1}{\e_{\vp}}   \, . 
\end{align}
Those relations~\eqref{AEM-relation},\eqref{AG-relation} are the main results of this subsection.

Verifying those expressions is straightforward. 
For example, by substituting $f_{\rm cov}$~\eqref{df-cov} and $\vg$~\eqref{fg-non-ana} with $\vzeta_{\EM}$ given by eq.~\eqref{zeta-EM},
we find eq.~\eqref{AEM-match} precisely equals to the spatial components of ${\bm\sA}_{\EM}$ in eq.~\eqref{A-field-int} 
\begin{align}
\label{vA-EM-field}
\bm{\sA}_{\EM}
 &= 
 \frac{\vv\cdot \vq}{v\cdot Q}f'_0
 \left[-
 \frac{(\vB-\vv\times \vE)}{\e_{\vp}}
 \right]- \frac{f_0}{\e_{\vp}^2}(\vv\times\vE) \, ,
\end{align}
We have utilized the relation
\begin{align}
    \vv\times {\bm\pd}f_{\rm cov}
    = \frac{\vv \cdot q}{v\cdot Q}(\vv \times\vE-\vv^2 \vB_{\perp})f'_0+ \vv^2\vB_{\perp} f'_0\, \,  . \no
\end{align}
which follows from the identity (using Faraday's law)
\begin{align}
 \vv\times\vpd(\vE\cdot \vv)
 = (\vv \cdot \vpd)(\vv \times\vE-\vv^2 \vB_{\perp})
 +\vv^2(v\cdot\pd)\vB_{\perp} \, . \no 
\end{align}
Since $\sA_{\EM}$ is expressed in terms of field strength, we use the covariant distribution $f_{\rm cov}$ in $\sA_{\rm con}$ (see eq.~\eqref{AEM-relation}). 
If we had used $\d f$~\eqref{fg-non-ana} with $\d \e_{\EM}$ in the evaluation of $\sA_{\rm con}$, 
the analytic piece in eq.~\eqref{AEM-relation} would look more complicated. 
The verification for the gravity case is analogous. 
Note, if we included the self-energy modification of Fermions in the field theory calculation, the particle/spin energy $(\d\e,\vzeta)$ should become a functional of the dynamical field, analogous to the case of Landau Fermi liquid.

We note that, in the linearized regime, one has certain freedom in choosing the form of the "spin-energy" $\vzeta$.
For instance, the dependence of $\vzeta_{\EM}$ on field strength can generally be expressed as ${\bm \zeta}= d_1\vB +d_2 \vB_{\perp}+ d_{3}(\vv\times \vE)$. 
In the expression~\eqref{AEM-match}, we have opted for $\vzeta\propto \vB$ (i,e. $d_2=d_3=0$). 
This choice is motivated by the physical consideration that the distribution of spin tends to align with the direction of the magnetic field, which can be achieved dynamically with \(\vzeta \propto \vB\), since \(\vg \propto \vzeta\) (see eq.~\eqref{fg-non-ana}).
\footnote{
The overall "$-$" sign found in $\vzeta_{\EM}$ is related to the overall sign convention for the gauge transformation~\eqref{U1}. 
}. 
Our choice for the form of \(\vzeta_{G}\) is similarly motivated by the parallels between \(\sA_{\EM}\) and \(\sA_{G}\). 
With a different selection of \(d\) values, one could still ensure that the full axial Wigner function remains unchanged in the linearized regime by adjusting the corresponding \(c\) values.
This must happen since \(\vg\) linearly depends on \(\zeta\) through eq.~\eqref{fg-non-ana}. 
Note in the full kinetic equation~\eqref{s-eq}, 
the non-linear term $\vg\times \vzeta$ is sensitive to the choice of $\vzeta$.
This indicates that we may analyze higher point functions to constrain the form of $\vzeta$, which we defer to future work.

The fact that the non-analytic part of $\sA$ can be expressed in terms of the dynamical field we propose is highly remarkable. 
This serves as a non-trivial test that all relevant slow degrees of freedom (d.o.fs) are properly identified.

We now turn to interpret the explicit expression for the constitutive relation obtained by substituting eq.~\eqref{cs} into eq.~\eqref{A-con-field}:
\begin{align}
\label{A-con-field}
{\bm \sA}_{\con}[\vg, f]=
\vg_{\parallel}+(1-\vv^2)\vg_{\perp}+\frac{1}{\e_{\vp}}\vv\times {\bm\pd}f\, ,
\qquad
\sA^0_{\rm con}[\vg, f]=\vv\cdot\vg\, . 
\end{align}
We observe that the prefactor for \(\vg_{\parallel}\) and \(\vg_{\perp}\) shows different dependence on \(\vv\). For relativistic fermions where \(\vv^2 = 1\), \(\vg_{\perp}\) does not contribute. This is anticipated because the spin of chiral fermions is aligned with the direction of their velocity.
The term proportional to \(\vv \times \vpd f\) resembles the magnetization current contribution to the vector Wigner function. Notably, for a particle at rest, \(\sA_{\rm con} = \vg\) as stated in equation \eqref{A-con-field}. 
This suggests that while \(\vg\) differs from the non-analytic part of \(\sA\), they may be related through a Lorentz boost. 
This connection warrants further investigation (perhaps along the lines of ref.~\cite{Ji:2020hii}).

In addition to $\sA_{\rm con}$, which is related to the dynamics of the distribution, there is also an analytic component that is interesting in its own right. 
For example, in ${\bm \sA}_{\EM}$~\eqref{AEM-relation}, we observe a term $\vv \times \vE$. This term manifests the spin Hall effect (SHE) (see Refs.~\cite{Hattori:2019ahi,Liu2021} for discussion in the context of QCD-like plasma, and ref.~\cite{Yamamoto:2023okm} for its application in the neutrino transport in core-collapse supernovae). 
Similarly, an analogous term appears in ${\bm \sA}_{\rm G}$~\eqref{AG-relation}, namely $\vv \times \vE^{h}$. In the homogeneous limit, one finds that $\vE^{h}_j = -(\partial_{t} h_{ij} v^j)/2$. In hydrodynamic limits, the response to the fluid shear stress $\sigma_{ij}$ can be mimicked by activating $\partial_{t} h_{ij} v^j$. In this sense, the shear-induced polarization~\cite{Fu:2021pok,Becattini_2021} can be heuristically thought of as an analogue of the SHE.

Since \(\sA_{\rm G}\) is invariant under linear diffeomorphism, we now show that the expression for \(\sA_{\rm G}\) appears simpler when expressed in a diffeomorphism-covariant form. In particular, the analytic component should vanish at first order in the gradient, as the lowest-order diffeomorphism-covariant terms are the Riemann curvature tensors \(R^{\mu\nu}_{\ \ \rho\a} \sim \mathcal{O}(\partial^2)\).

Let us begin by rewriting eq.~\eqref{A-metric} in a manifestly diffeomorphism-invariant form:
\begin{align}
\mathcal{A}^{\mu} = \frac{i}{2} \frac{1}{v \cdot Q} \tilde{R}^{\mu}_{\ \rho\sigma\nu} v^{\rho} v^{\sigma} v_{\perp}^{\nu} (-f_0'),
\end{align}
or equivalently,
\begin{align}
\label{A-G-R}
\sA^i_{\rm G} = \frac{i}{v \cdot Q} f_0' \left[\frac{1}{2} \left( \tilde{R}^{0i}_{\ \ 0l} - \epsilon^{im}_{\,\,\,\, \,\, n} v_{m} R^{0n}_{\ \ 0l} \right) v^{l}\right], \quad 
\sA^{0}_{\rm G} = \bm{\sA}_{\rm G} \cdot \vv.
\end{align}
Noting the similarity between eq.~\eqref{A-G-R} and the first term in eq.~\eqref{vA-EM-field}, we are not surprised by the relationship we obtain through explicit verification:
\begin{align}
\bm{\sA}_{\rm G} = {\bm\sA}_{\rm con}[\vg_{\rm cov}, f_{\rm cov}]\,.
\end{align}
Here, \({\bm\sA}_{\rm con}\) takes the same form as in eq.~\eqref{A-con-field}, except that it replaces \(\partial_{\mu}\) with the covariant derivative \(D_{\mu}\) as indicated in eq.~\eqref{D-cov}.

Similar to the case for \(f_{\rm cov}\), the linearized equation for the covariant spin distribution \(\vg_{\rm cov}\) does not adhere to the Poisson bracket structure but is analogous to the equation for \(f_{\rm cov}\) in eq.~\eqref{fcov-eq}:
\begin{align}
(v \cdot D)g^i_{\rm cov} + \left(-\frac{1}{2} \tilde{R}^{0i}_{\ \ 0l} v^{l}\right) f_0' = 0\,.
\end{align}

To end this section, we mention that other components of the Wigner function in the linear response regime can be expressed in terms of the dynamical variable $(f,\vg)$, following similar steps above.

\section{Discussion and outlook}

In this paper, we presented the bottom-up formulation of spin kinetic theory (SKT). 
This approach exhibits several qualitative distinctions relative to widely used top-down methods.

First, the choice of dynamical variables differs.
In this work, we introduce \((f,\vg)\) to parametrize the dynamics of the spin-averaged and the spin-dependent distribution. 
In contrast, the vector and axial Wigner functions, \(\sV\) and \(\sA\), are commonly employed as the dynamical variables in the top-down approach (e.g., see Refs. \cite{Hattori:2019ahi, Weickgenannt:2020aaf}).
If we were writing down the E.o.M. full Wigner function, such as \(\sA\), it would look more complicated (see, for example, eq. 27 in ref. \cite{Hattori:2019ahi}) compared to the E.o.M. for \((f, \vg)\). 
This complexity does not stem from the description of the dynamics itself, but rather from the presence of non-dynamical terms in \(\sV\) and \(\sA\). 
Therefore, it is not surprising that the E.o.M. is generally simplified in the present approach while still adequately describing the dynamics.

Second, we employ the spin-dependent Poisson Bracket (P.B.)~\cite{Delacretaz:2022ocm} to formulate the E.o.M. in the collisionless regime, while E.o.M. in the top-down approaches resort to the Kadanoff-Baym equation.
We observe that spin-dependent P.B. implies an intrinsic non-linear E.o.M, with non-linear term (such as $\bm\zeta\times \vg$ in eq.~\eqref{eq-fg}) completely fixed by P.B. algebra.
This indicates that the P.B. can be effectively utilized to simplify the derivation of the kinetic equation and to highlight its algebraic structures.
Additionally, we hope that the P.B. algebra is still useful in the non-perturbative regime of QCD plasma.
In fact, the formulations presented in Refs. \cite{Delacretaz:2022ocm,Else_2023,Huang:2024uap} are motivated by the desire to describe strongly coupled non-Fermi liquids.

Third, the bottom-up framework does not begin with a specific microscopic theory, making the generality of SKT transparent. 
We have already seen that the P.B. is purely determined by algebra. 
Furthermore, the functional dependence on the dynamical variables in the constitutive relation comes from general considerations. 
The only microscopic input is encoded in the "transport coefficients", similar to the formulation of hydrodynamics. 
We have verified that the dependence of $\sA$ on dynamical variable $(f,\vg)$ does not depend on whether these distributions are induced by electromagnetic fields or metric perturbations.

We stress that the constitutive relation in the kinetic theory differs from hydrodynamics in that it is applicable even in far-from-equilibrium conditions.
This is particularly relevant given the recent observations of hyperon polarization in p-Pb collisions, even in events with very low multiplicity, as reported by the CMS collaboration.~\cite {CMS:2025nqr}.
When local equilibrium is reached, $\sA$ can be expressed in terms of the hydrodynamic gradient~\cite{Liu:2021uhn}. 
The resulting expression is used in most phenomenological analysis of spin effects, although in low-multiplicity collision events, it is unclear whether local equilibrium is achieved.
In contrast, the constitutive relation in terms of non-equilibrium distributions may still provide valuable insights. Notably, the presence of the analogous magnetization current term as discussed earlier
\begin{align}
\label{MC}
    {\bm\sA}(x,\vv) \propto \vv \times \nabla f\, 
\end{align}
indicates that an inhomogeneous, spin-independent far-from-equilibrium distribution can polarize spin in phase space. 
A quantitative analysis based on~\eqref{MC} may help interpret the intriguing results mentioned above.

Our study exercises the power of the bottom-up approach in a context where we have made several simplifications. 
We expect, though, that none of these simplifications reflects the limitation of the bottom-up approach. It would be valuable to establish the matching to field-theoretical calculations beyond the lowest order in perturbation theory, along the lines of ref.~\cite{Chen2016}. 
We also wish to implement the Lorentz transformation in addition to the spacetime translation in the action. 
This would enable us to express the phase-space current associated with angular momentum and thereby investigate the mechanisms that transfer orbital angular momentum to quark spins. 
Another area where the extension of spin-dependent P.B. could be valuable is in ultra-dense QCD matter, where quarks may form Cooper pairs and where spin is intertwined with color (see ref.~\cite{Sogabe:2024yfl} for the topological structure resulting from spin-color locking in one-flavor QCD).

We have presented the Schwinger-Keldysh action for the collisionless SKT, which, to our knowledge, is new in the literature. 
Starting with the action principle may help clarify various theoretical subtleties in the description of spin, such as those related to the pseudo-gauge transformation and the choice of spin frame. 
The formulation of a non-hydrodynamic EFT based on symmetry and the action principle has recently attracted interest~\cite{An:2025ils}.
The spin distribution presents a simple yet non-trivial example of a non-hydrodynamic degree of freedom with non-Abelian symmetry ($SU(2)$) that allows for an action formulation.

The proton exhibits a complex spin structure when its angular momentum is decomposed into the contributions from partons, which are in a confined state (see ref. \cite{Ji:2020ena}). 
This raises a natural question: does the deconfined hot QCD plasma also feature a non-trivial spin structure in an inhomogeneous environment? 
If so, can this structure be explained through quark-gluon interactions?
We hope that the developments presented here will eventually advance our understanding of these issues.

\acknowledgments
We thank Jian-hua Gao, Jin Hu, Jeremy Hansen, Xiao-Yang Huang, Shu Lin, Shi Pu, Misha Stephahov, and Ho-Ung Yee for helpful discussions/comments.
This work is supported partly by NSFC under Grant No.12575150 and by CUHK-Shenzhen University Development Fund under the Grant No.UDF01003791 (YY) and partly by the U.S. Department of
Energy, Office of Science, Office of Nuclear Physics, grant No. DEFG0201ER41195 (Z.L.M).


\bibliographystyle{JHEP.bst}
\bibliography{ref}

\providecommand{\href}[2]{#2}\begingroup\raggedright\begin{thebibliography}{10}

\bibitem{STAR:2017ckg}
{\scshape STAR} collaboration, \emph{{Global $\Lambda$ hyperon polarization in
  nuclear collisions: evidence for the most vortical fluid}},
  \href{https://doi.org/10.1038/nature23004}{\emph{Nature} {\bfseries 548}
  (2017) 62}.

\bibitem{STAR:2019erd}
{\scshape STAR} collaboration, \emph{{Polarization of $\Lambda$
  ($\bar{\Lambda}$) hyperons along the beam direction in Au+Au collisions at
  $\sqrt{s_{_{NN}}}$ = 200 GeV}},
  \href{https://doi.org/10.1103/PhysRevLett.123.132301}{\emph{Phys. Rev. Lett.}
  {\bfseries 123} (2019) 132301}
  [\href{https://arxiv.org/abs/1905.11917}{{\ttfamily 1905.11917}}].

\bibitem{STAR:2022fan}
{\scshape STAR} collaboration, \emph{{Pattern of global spin alignment of
  \ensuremath{\phi} and K$^{*0}$ mesons in heavy-ion collisions}},
  \href{https://doi.org/10.1038/s41586-022-05557-5}{\emph{Nature} {\bfseries
  614} (2023) 244} [\href{https://arxiv.org/abs/2204.02302}{{\ttfamily
  2204.02302}}].

\bibitem{Sheng:2022wsy}
X.-L.~Sheng, L.~Oliva, Z.-T.~Liang, Q.~Wang and X.-N.~Wang, \emph{{Spin
  Alignment of Vector Mesons in Heavy-Ion Collisions}},
  \href{https://doi.org/10.1103/PhysRevLett.131.042304}{\emph{Phys. Rev. Lett.}
  {\bfseries 131} (2023) 042304}
  [\href{https://arxiv.org/abs/2205.15689}{{\ttfamily 2205.15689}}].

\bibitem{Liang:2004ph}
Z.-T.~Liang and X.-N.~Wang, \emph{{Globally polarized quark-gluon plasma in
  non-central A+A collisions}},
  \href{https://doi.org/10.1103/PhysRevLett.94.102301}{\emph{Phys. Rev. Lett.}
  {\bfseries 94} (2005) 102301}
  [\href{https://arxiv.org/abs/nucl-th/0410079}{{\ttfamily nucl-th/0410079}}].

\bibitem{Becattini:2013fla}
F.~Becattini, V.~Chandra, L.~Del~Zanna and E.~Grossi, \emph{{Relativistic
  distribution function for particles with spin at local thermodynamical
  equilibrium}}, \href{https://doi.org/10.1016/j.aop.2013.07.004}{\emph{Annals
  Phys.} {\bfseries 338} (2013) 32}.

\bibitem{Liu:2021uhn}
S.Y.F.~Liu and Y.~Yin, \emph{{Spin polarization induced by the hydrodynamic
  gradients}}, \href{https://doi.org/10.1007/JHEP07(2021)188}{\emph{JHEP}
  {\bfseries 07} (2021) 188}
  [\href{https://arxiv.org/abs/2103.09200}{{\ttfamily 2103.09200}}].

\bibitem{Fu:2021pok}
B.~Fu, S.Y.F.~Liu, L.~Pang, H.~Song and Y.~Yin, \emph{{Shear-Induced Spin
  Polarization in Heavy-Ion Collisions}},
  \href{https://doi.org/10.1103/PhysRevLett.127.142301}{\emph{Phys. Rev. Lett.}
  {\bfseries 127} (2021) 142301}
  [\href{https://arxiv.org/abs/2103.10403}{{\ttfamily 2103.10403}}].

\bibitem{Becattini:2021suc}
F.~Becattini, M.~Buzzegoli and A.~Palermo, \emph{{Spin-thermal shear coupling
  in a relativistic fluid}},
  \href{https://doi.org/10.1016/j.physletb.2021.136519}{\emph{Phys. Lett. B}
  {\bfseries 820} (2021) 136519}
  [\href{https://arxiv.org/abs/2103.10917}{{\ttfamily 2103.10917}}].

\bibitem{Becattini:2021iol}
F.~Becattini, M.~Buzzegoli, G.~Inghirami, I.~Karpenko and A.~Palermo,
  \emph{{Local Polarization and Isothermal Local Equilibrium in Relativistic
  Heavy Ion Collisions}},
  \href{https://doi.org/10.1103/PhysRevLett.127.272302}{\emph{Phys. Rev. Lett.}
  {\bfseries 127} (2021) 272302}
  [\href{https://arxiv.org/abs/2103.14621}{{\ttfamily 2103.14621}}].

\bibitem{Liu:2020dxg}
S.Y.F.~Liu and Y.~Yin, \emph{{Spin Hall effect in heavy-ion collisions}},
  \href{https://doi.org/10.1103/PhysRevD.104.054043}{\emph{Phys. Rev. D}
  {\bfseries 104} (2021) 054043}
  [\href{https://arxiv.org/abs/2006.12421}{{\ttfamily 2006.12421}}].

\bibitem{Fu:2022myl}
B.~Fu, L.~Pang, H.~Song and Y.~Yin, \emph{{Signatures of the spin Hall effect
  in hot and dense QCD matter}},
  \href{https://arxiv.org/abs/2201.12970}{{\ttfamily 2201.12970}}.

\bibitem{Becattini:2020ngo}
F.~Becattini and M.A.~Lisa, \emph{{Polarization and Vorticity in the
  Quark\textendash{}Gluon Plasma}},
  \href{https://doi.org/10.1146/annurev-nucl-021920-095245}{\emph{Ann. Rev.
  Nucl. Part. Sci.} {\bfseries 70} (2020) 395}
  [\href{https://arxiv.org/abs/2003.03640}{{\ttfamily 2003.03640}}].

\bibitem{Becattini2022}
F.~Becattini, \emph{Spin and polarization: a new direction in relativistic
  heavy ion physics}, .

\bibitem{YinYi:2023msj}
Y.~Yi, \emph{{Quantum correlation between spin and motion in quantum
  chromodynamics matter}},
  \href{https://doi.org/10.7498/aps.72.20222458}{\emph{Acta Phys. Sin.}
  {\bfseries 72} (2023) 111201}.

\bibitem{Hattori:2019ahi}
K.~Hattori, Y.~Hidaka and D.-L.~Yang, \emph{{Axial Kinetic Theory and Spin
  Transport for Fermions with Arbitrary Mass}},
  \href{https://doi.org/10.1103/PhysRevD.100.096011}{\emph{Phys. Rev. D}
  {\bfseries 100} (2019) 096011}
  [\href{https://arxiv.org/abs/1903.01653}{{\ttfamily 1903.01653}}].

\bibitem{Weickgenannt:2019dks}
N.~Weickgenannt, X.-L.~Sheng, E.~Speranza, Q.~Wang and D.H.~Rischke,
  \emph{{Kinetic theory for massive spin-1/2 particles from the Wigner-function
  formalism}}, \href{https://doi.org/10.1103/PhysRevD.100.056018}{\emph{Phys.
  Rev. D} {\bfseries 100} (2019) 056018}
  [\href{https://arxiv.org/abs/1902.06513}{{\ttfamily 1902.06513}}].

\bibitem{Gao:2019znl}
J.-H.~Gao and Z.-T.~Liang, \emph{{Relativistic Quantum Kinetic Theory for
  Massive Fermions and Spin Effects}},
  \href{https://doi.org/10.1103/PhysRevD.100.056021}{\emph{Phys. Rev. D}
  {\bfseries 100} (2019) 056021}
  [\href{https://arxiv.org/abs/1902.06510}{{\ttfamily 1902.06510}}].

\bibitem{Weickgenannt:2020aaf}
N.~Weickgenannt, E.~Speranza, X.-l.~Sheng, Q.~Wang and D.H.~Rischke,
  \emph{{Generating Spin Polarization from Vorticity through Nonlocal
  Collisions}},
  \href{https://doi.org/10.1103/PhysRevLett.127.052301}{\emph{Phys. Rev. Lett.}
  {\bfseries 127} (2021) 052301}
  [\href{https://arxiv.org/abs/2005.01506}{{\ttfamily 2005.01506}}].

\bibitem{Weickgenannt:2021cuo}
N.~Weickgenannt, E.~Speranza, X.-l.~Sheng, Q.~Wang and D.H.~Rischke,
  \emph{{Derivation of the nonlocal collision term in the relativistic
  Boltzmann equation for massive spin-1/2 particles from quantum field
  theory}}, \href{https://doi.org/10.1103/PhysRevD.104.016022}{\emph{Phys. Rev.
  D} {\bfseries 104} (2021) 016022}
  [\href{https://arxiv.org/abs/2103.04896}{{\ttfamily 2103.04896}}].

\bibitem{Hayata2020}
T.~Hayata, Y.~Hidaka and K.~Mameda, \emph{{Second order chiral kinetic theory
  under gravity and antiparallel charge-energy flow}},
  \href{https://arxiv.org/abs/2012.12494}{{\ttfamily 2012.12494}}.

\bibitem{Sheng:2021kfc}
X.-L.~Sheng, N.~Weickgenannt, E.~Speranza, D.H.~Rischke and Q.~Wang,
  \emph{{From Kadanoff-Baym to Boltzmann equations for massive spin-1/2
  fermions}}, \href{https://doi.org/10.1103/PhysRevD.104.016029}{\emph{Phys.
  Rev. D} {\bfseries 104} (2021) 016029}
  [\href{https://arxiv.org/abs/2103.10636}{{\ttfamily 2103.10636}}].

\bibitem{Yang:2020hri}
D.-L.~Yang, K.~Hattori and Y.~Hidaka, \emph{{Effective quantum kinetic theory
  for spin transport of fermions with collsional effects}},
  \href{https://doi.org/10.1007/JHEP07(2020)070}{\emph{JHEP} {\bfseries 07}
  (2020) 070} [\href{https://arxiv.org/abs/2002.02612}{{\ttfamily
  2002.02612}}].

\bibitem{Hidaka:2022dmn}
Y.~Hidaka, S.~Pu, Q.~Wang and D.-L.~Yang, \emph{{Foundations and applications
  of quantum kinetic theory}},
  \href{https://doi.org/10.1016/j.ppnp.2022.103989}{\emph{Prog. Part. Nucl.
  Phys.} {\bfseries 127} (2022) 103989}
  [\href{https://arxiv.org/abs/2201.07644}{{\ttfamily 2201.07644}}].

\bibitem{Son:2012wh}
D.T.~Son and N.~Yamamoto, \emph{{Berry Curvature, Triangle Anomalies, and the
  Chiral Magnetic Effect in Fermi Liquids}},
  \href{https://doi.org/10.1103/PhysRevLett.109.181602}{\emph{Phys. Rev. Lett.}
  {\bfseries 109} (2012) 181602}
  [\href{https://arxiv.org/abs/1203.2697}{{\ttfamily 1203.2697}}].

\bibitem{Stephanov:2012ki}
M.A.~Stephanov and Y.~Yin, \emph{{Chiral Kinetic Theory}},
  \href{https://doi.org/10.1103/PhysRevLett.109.162001}{\emph{Phys. Rev. Lett.}
  {\bfseries 109} (2012) 162001}
  [\href{https://arxiv.org/abs/1207.0747}{{\ttfamily 1207.0747}}].

\bibitem{Son:2012zy}
D.T.~Son and N.~Yamamoto, \emph{{Kinetic theory with Berry curvature from
  quantum field theories}},
  \href{https://doi.org/10.1103/PhysRevD.87.085016}{\emph{Phys. Rev. D}
  {\bfseries 87} (2013) 085016}
  [\href{https://arxiv.org/abs/1210.8158}{{\ttfamily 1210.8158}}].

\bibitem{Chen:2012ca}
J.-W.~Chen, S.~Pu, Q.~Wang and X.-N.~Wang, \emph{{Berry Curvature and
  Four-Dimensional Monopoles in the Relativistic Chiral Kinetic Equation}},
  \href{https://doi.org/10.1103/PhysRevLett.110.262301}{\emph{Phys. Rev. Lett.}
  {\bfseries 110} (2013) 262301}
  [\href{https://arxiv.org/abs/1210.8312}{{\ttfamily 1210.8312}}].

\bibitem{Gao:2012ix}
J.-H.~Gao, Z.-T.~Liang, S.~Pu, Q.~Wang and X.-N.~Wang, \emph{{Chiral Anomaly
  and Local Polarization Effect from Quantum Kinetic Approach}},
  \href{https://doi.org/10.1103/PhysRevLett.109.232301}{\emph{Phys. Rev. Lett.}
  {\bfseries 109} (2012) 232301}
  [\href{https://arxiv.org/abs/1203.0725}{{\ttfamily 1203.0725}}].

\bibitem{Chen:2013iga}
J.-W.~Chen, J.-y.~Pang, S.~Pu and Q.~Wang, \emph{{Kinetic equations for massive
  Dirac fermions in electromagnetic field with non-Abelian Berry phase}},
  \href{https://doi.org/10.1103/PhysRevD.89.094003}{\emph{Phys. Rev. D}
  {\bfseries 89} (2014) 094003}
  [\href{https://arxiv.org/abs/1312.2032}{{\ttfamily 1312.2032}}].

\bibitem{Manuel:2013zaa}
C.~Manuel and J.M.~Torres-Rincon, \emph{{Kinetic theory of chiral relativistic
  plasmas and energy density of their gauge collective excitations}},
  \href{https://doi.org/10.1103/PhysRevD.89.096002}{\emph{Phys. Rev. D}
  {\bfseries 89} (2014) 096002}
  [\href{https://arxiv.org/abs/1312.1158}{{\ttfamily 1312.1158}}].

\bibitem{Manuel:2014dza}
C.~Manuel and J.M.~Torres-Rincon, \emph{{Chiral transport equation from the
  quantum Dirac Hamiltonian and the on-shell effective field theory}},
  \href{https://doi.org/10.1103/PhysRevD.90.076007}{\emph{Phys. Rev. D}
  {\bfseries 90} (2014) 076007}
  [\href{https://arxiv.org/abs/1404.6409}{{\ttfamily 1404.6409}}].

\bibitem{Chen:2014cla}
J.-Y.~Chen, D.T.~Son, M.A.~Stephanov, H.-U.~Yee and Y.~Yin, \emph{{Lorentz
  Invariance in Chiral Kinetic Theory}},
  \href{https://doi.org/10.1103/PhysRevLett.113.182302}{\emph{Phys. Rev. Lett.}
  {\bfseries 113} (2014) 182302}
  [\href{https://arxiv.org/abs/1404.5963}{{\ttfamily 1404.5963}}].

\bibitem{Chen:2015gta}
J.-Y.~Chen, D.T.~Son and M.A.~Stephanov, \emph{{Collisions in Chiral Kinetic
  Theory}}, \href{https://doi.org/10.1103/PhysRevLett.115.021601}{\emph{Phys.
  Rev. Lett.} {\bfseries 115} (2015) 021601}
  [\href{https://arxiv.org/abs/1502.06966}{{\ttfamily 1502.06966}}].

\bibitem{Hidaka:2016yjf}
Y.~Hidaka, S.~Pu and D.-L.~Yang, \emph{{Relativistic Chiral Kinetic Theory from
  Quantum Field Theories}},
  \href{https://doi.org/10.1103/PhysRevD.95.091901}{\emph{Phys. Rev. D}
  {\bfseries 95} (2017) 091901}
  [\href{https://arxiv.org/abs/1612.04630}{{\ttfamily 1612.04630}}].

\bibitem{Huang:2018wdl}
A.~Huang, S.~Shi, Y.~Jiang, J.~Liao and P.~Zhuang, \emph{{Complete and
  Consistent Chiral Transport from Wigner Function Formalism}},
  \href{https://doi.org/10.1103/PhysRevD.98.036010}{\emph{Phys. Rev. D}
  {\bfseries 98} (2018) 036010}
  [\href{https://arxiv.org/abs/1801.03640}{{\ttfamily 1801.03640}}].

\bibitem{Gao:2018wmr}
J.-H.~Gao, Z.-T.~Liang, Q.~Wang and X.-N.~Wang, \emph{{Disentangling covariant
  Wigner functions for chiral fermions}},
  \href{https://doi.org/10.1103/PhysRevD.98.036019}{\emph{Phys. Rev. D}
  {\bfseries 98} (2018) 036019}
  [\href{https://arxiv.org/abs/1802.06216}{{\ttfamily 1802.06216}}].

\bibitem{Ma:2022ins}
S.-X.~Ma and J.-H.~Gao, \emph{{Generalized chiral kinetic equations}},
  \href{https://doi.org/10.1016/j.physletb.2023.138100}{\emph{Phys. Lett. B}
  {\bfseries 844} (2023) 138100}
  [\href{https://arxiv.org/abs/2209.10737}{{\ttfamily 2209.10737}}].

\bibitem{lifshitz1981kinetics}
E.M.~Lifshitz and L.P.~Pitaevskii, \emph{Physical Kinetics}, vol.~10 of
  \emph{Course of Theoretical Physics}, Pergamon Press, Oxford (1981).

\bibitem{Mueller:2002gd}
A.H.~Mueller and D.T.~Son, \emph{{On the Equivalence between the Boltzmann
  equation and classical field theory at large occupation numbers}},
  \href{https://doi.org/10.1016/j.physletb.2003.12.047}{\emph{Phys. Lett. B}
  {\bfseries 582} (2004) 279}
  [\href{https://arxiv.org/abs/hep-ph/0212198}{{\ttfamily hep-ph/0212198}}].

\bibitem{Chen2016}
J.-Y.~Chen and D.T.~Son, \emph{{Berry Fermi Liquid Theory}},
  \href{https://arxiv.org/abs/1604.07857}{{\ttfamily 1604.07857}}.

\bibitem{Blaizot:1993zk}
J.P.~Blaizot and E.~Iancu, \emph{{Kinetic equations for long wavelength
  excitations of the quark - gluon plasma}},
  \href{https://doi.org/10.1103/PhysRevLett.70.3376}{\emph{Phys. Rev. Lett.}
  {\bfseries 70} (1993) 3376}
  [\href{https://arxiv.org/abs/hep-ph/9301236}{{\ttfamily hep-ph/9301236}}].

\bibitem{abrikosov1975methods}
A.A.~Abrikosov, L.P.~Gorkov and I.E.~Dzyaloshinski, \emph{Methods of Quantum
  Field Theory in Statistical Physics}, Dover Publications, New York,
  revised~ed. (1975).

\bibitem{Delacretaz:2022ocm}
L.V.~Delacretaz, Y.-H.~Du, U.~Mehta and D.T.~Son, \emph{{Nonlinear bosonization
  of Fermi surfaces: The method of coadjoint orbits}},
  \href{https://doi.org/10.1103/PhysRevResearch.4.033131}{\emph{Phys. Rev.
  Res.} {\bfseries 4} (2022) 033131}
  [\href{https://arxiv.org/abs/2203.05004}{{\ttfamily 2203.05004}}].

\bibitem{Ghiglieri:2020dpq}
J.~Ghiglieri, A.~Kurkela, M.~Strickland and A.~Vuorinen, \emph{{Perturbative
  Thermal QCD: Formalism and Applications}},
  \href{https://doi.org/10.1016/j.physrep.2020.07.004}{\emph{Phys. Rept.}
  {\bfseries 880} (2020) 1} [\href{https://arxiv.org/abs/2002.10188}{{\ttfamily
  2002.10188}}].

\bibitem{Liu:2018kfw}
H.~Liu and P.~Glorioso, \emph{{Lectures on non-equilibrium effective field
  theories and fluctuating hydrodynamics}},
  \href{https://doi.org/10.22323/1.305.0008}{\emph{PoS} {\bfseries TASI2017}
  (2018) 008} [\href{https://arxiv.org/abs/1805.09331}{{\ttfamily
  1805.09331}}].

\bibitem{Yang:2024sfp}
S.-Z.~Yang, J.-H.~Gao and S.~Pu, \emph{{Corrections from spacetime dependent
  electromagnetic fields to Wigner functions and spin polarization}},
  \href{https://doi.org/10.1103/PhysRevD.111.036013}{\emph{Phys. Rev. D}
  {\bfseries 111} (2025) 036013}
  [\href{https://arxiv.org/abs/2409.00456}{{\ttfamily 2409.00456}}].

\bibitem{Chou:1984es}
K.-c.~Chou, Z.-b.~Su, B.-l.~Hao and L.~Yu, \emph{{Equilibrium and
  Nonequilibrium Formalisms Made Unified}},
  \href{https://doi.org/10.1016/0370-1573(85)90136-X}{\emph{Phys. Rept.}
  {\bfseries 118} (1985) 1}.

\bibitem{Crossley:2015evo}
M.~Crossley, P.~Glorioso and H.~Liu, \emph{{Effective field theory of
  dissipative fluids}},
  \href{https://doi.org/10.1007/JHEP09(2017)095}{\emph{JHEP} {\bfseries 09}
  (2017) 095} [\href{https://arxiv.org/abs/1511.03646}{{\ttfamily
  1511.03646}}].

\bibitem{Martin:1973zz}
P.C.~Martin, E.D.~Siggia and H.A.~Rose, \emph{{Statistical Dynamics of
  Classical Systems}},
  \href{https://doi.org/10.1103/PhysRevA.8.423}{\emph{Phys. Rev. A} {\bfseries
  8} (1973) 423}.

\bibitem{Delacretaz:2025ifh}
L.V.~Delacr{\'e}taz, S.D.~Chowdhury and U.~Mehta, \emph{{Symmetry and causality
  constraints on Fermi liquids}},
  \href{https://arxiv.org/abs/2501.02073}{{\ttfamily 2501.02073}}.

\bibitem{Heinz:1983nx}
U.W.~Heinz, \emph{{Kinetic Theory for Nonabelian Plasmas}},
  \href{https://doi.org/10.1103/PhysRevLett.51.351}{\emph{Phys. Rev. Lett.}
  {\bfseries 51} (1983) 351}.

\bibitem{Fonarev:1993ht}
O.A.~Fonarev, \emph{{Wigner function and quantum kinetic theory in curved
  space-time and external fields}},
  \href{https://doi.org/10.1063/1.530542}{\emph{J. Math. Phys.} {\bfseries 35}
  (1994) 2105} [\href{https://arxiv.org/abs/gr-qc/9309005}{{\ttfamily
  gr-qc/9309005}}].

\bibitem{PhysRevD.37.2901}
E.~Calzetta, S.~Habib and B.L.~Hu, \emph{Quantum kinetic field theory in curved
  spacetime: Covariant wigner function and liouville-vlasov equations},
  \href{https://doi.org/10.1103/PhysRevD.37.2901}{\emph{Phys. Rev. D}
  {\bfseries 37} (1988) 2901}.

\bibitem{Antonsen:1997dc}
F.~Antonsen, \emph{{A New nonperturbative approach to quantum theory in curved
  space-time using the Wigner function}},
  \href{https://doi.org/10.1103/PhysRevD.56.920}{\emph{Phys. Rev. D} {\bfseries
  56} (1997) 920} [\href{https://arxiv.org/abs/hep-th/9701182}{{\ttfamily
  hep-th/9701182}}].

\bibitem{Liu2018}
Y.-C.~Liu, L.-L.~Gao, K.~Mameda and X.-G.~Huang, \emph{{Chiral kinetic theory
  in curved spacetime}},  \href{https://arxiv.org/abs/1812.10127}{{\ttfamily
  1812.10127}}.

\bibitem{Lin:2024svh}
S.~Lin and J.~Tian, \emph{{Spin polarized quasiparticle in an off-equilibrium
  medium}}, \href{https://doi.org/10.1103/f92k-667g}{\emph{Phys. Rev. D}
  {\bfseries 112} (2025) 034005}
  [\href{https://arxiv.org/abs/2410.22935}{{\ttfamily 2410.22935}}].

\bibitem{mo2026bottom}
Z.~Mo, \emph{``bottom-up'' spin kinetic theory and its application to
  gravitational response},  Master's thesis, University of Science and
  Technology of China, 2026.

\bibitem{Caron-Huot:2007cma}
S.~Caron-Huot, \emph{{Hard thermal loops in the real-time formalism}},
  \href{https://doi.org/10.1088/1126-6708/2009/04/004}{\emph{JHEP} {\bfseries
  04} (2009) 004} [\href{https://arxiv.org/abs/0710.5726}{{\ttfamily
  0710.5726}}].

\bibitem{Gao:2020ksg}
H.~Gao, Z.~Mo and S.~Lin, \emph{{Photon self-energy in a magnetized chiral
  plasma from kinetic theory}},
  \href{https://doi.org/10.1103/PhysRevD.102.014011}{\emph{Phys. Rev. D}
  {\bfseries 102} (2020) 014011}
  [\href{https://arxiv.org/abs/2002.07959}{{\ttfamily 2002.07959}}].

\bibitem{Ji:2020hii}
X.~Ji and F.~Yuan, \emph{{Transverse spin sum rule of the proton}},
  \href{https://doi.org/10.1016/j.physletb.2020.135786}{\emph{Phys. Lett. B}
  {\bfseries 810} (2020) 135786}
  [\href{https://arxiv.org/abs/2008.04349}{{\ttfamily 2008.04349}}].

\bibitem{Liu2021}
S.Y.~Liu and Y.~Yin, \emph{{Spin polarization induced by the hydrodynamic
  gradients}}, \href{https://doi.org/10.1007/JHEP07(2021)188}{\emph{Journal of
  High Energy Physics} {\bfseries 2021} (2021) }
  [\href{https://arxiv.org/abs/2103.09200}{{\ttfamily 2103.09200}}].

\bibitem{Yamamoto:2023okm}
N.~Yamamoto and D.-L.~Yang, \emph{{Chiral kinetic theory with self-energy
  corrections and neutrino spin Hall effect}},
  \href{https://doi.org/10.1103/PhysRevD.109.056010}{\emph{Phys. Rev. D}
  {\bfseries 109} (2024) 056010}
  [\href{https://arxiv.org/abs/2308.08257}{{\ttfamily 2308.08257}}].

\bibitem{Becattini_2021}
F.~Becattini, M.~Buzzegoli and A.~Palermo, \emph{Spin-thermal shear coupling in
  a relativistic fluid},
  \href{https://doi.org/10.1016/j.physletb.2021.136519}{\emph{Physics Letters
  B} {\bfseries 820} (2021) 136519}.

\bibitem{Else_2023}
D.V.~Else, \emph{Collisionless dynamics of general non-fermi liquids from
  hydrodynamics of emergent conserved quantities},
  \href{https://doi.org/10.1103/physrevb.108.045107}{\emph{Physical Review B}
  {\bfseries 108} (2023) }.

\bibitem{Huang:2024uap}
X.~Huang, A.~Lucas, U.~Mehta and M.~Qi, \emph{{Effective field theory for
  ersatz Fermi liquids}},
  \href{https://doi.org/10.1103/PhysRevB.110.035102}{\emph{Phys. Rev. B}
  {\bfseries 110} (2024) 035102}
  [\href{https://arxiv.org/abs/2402.14066}{{\ttfamily 2402.14066}}].

\bibitem{CMS:2025nqr}
{\scshape CMS} collaboration, \emph{{Observation of {\ensuremath{\Lambda}}
  Hyperon Local Polarization in p-Pb Collisions at sNN=8.16{\,}{\,}TeV}},
  \href{https://doi.org/10.1103/6ywq-gm61}{\emph{Phys. Rev. Lett.} {\bfseries
  135} (2025) 132301} [\href{https://arxiv.org/abs/2502.07898}{{\ttfamily
  2502.07898}}].

\bibitem{Sogabe:2024yfl}
N.~Sogabe and Y.~Yin, \emph{{Berry Curvature and Spin-One Color
  Superconductivity}},
  \href{https://doi.org/10.1103/PhysRevLett.134.171903}{\emph{Phys. Rev. Lett.}
  {\bfseries 134} (2025) 171903}
  [\href{https://arxiv.org/abs/2411.08005}{{\ttfamily 2411.08005}}].

\bibitem{An:2025ils}
X.~An, R.~Brants, M.P.~Heller and Y.~Yin, \emph{{Building far-from-equilibrium
  effective field theories using shift symmetries}},
  \href{https://arxiv.org/abs/2511.11555}{{\ttfamily 2511.11555}}.

\bibitem{Ji:2020ena}
X.~Ji, F.~Yuan and Y.~Zhao, \emph{{What we know and what we
  don{\textquoteright}t know about the proton spin after 30 years}},
  \href{https://doi.org/10.1038/s42254-020-00248-4}{\emph{Nature Rev. Phys.}
  {\bfseries 3} (2021) 27} [\href{https://arxiv.org/abs/2009.01291}{{\ttfamily
  2009.01291}}].

\end{thebibliography}\endgroup

\end{document}